\documentclass[twocolumn,numberedappendix,iop,floatfix]{openjournal} 
\usepackage[utf8]{inputenc}
\usepackage[T1]{fontenc}

\usepackage{bm}
\usepackage{physics}

\usepackage{multirow}
\usepackage{cancel}

\usepackage[colorlinks,linkcolor=blue,citecolor=blue,urlcolor=blue]{hyperref}

\usepackage[nameinlink,noabbrev]{cleveref}
\crefname{equation}{Eq.}{Eqs.}
\crefname{figure}{Fig.}{Figs.}
\crefname{table}{Table}{Tables}
\crefname{section}{Section}{Sections}
\crefname{appendix}{Appendix}{Appendices} 
\Crefname{equation}{Equation}{Equations}
\Crefname{figure}{Figure}{Figures}

\usepackage{etoolbox}
\makeatletter
\patchcmd\H@refstepcounter{\protected@edef}{\protected@xdef}{}{\fail} 
\pretocmd\@sect{\def\@currentcounter{#1}}{}{\fail} 
\makeatother

\usepackage{orcidlink}

\newcommand{\ihMpc}{h^{-1}{\rm Mpc}}
\newcommand{\ihGpc}{h^{-1}{\rm Gpc}}

\newcommand{\safesc}[1]{\texorpdfstring{{\sc #1}}{#1}}

\newcommand{\pycorr}{{\sc pycorr}}

\newcommand{\abacus}{\safesc{Abacus}}
\newcommand{\abacussummit}{\safesc{AbacusSummit}}
\newcommand{\compaso}{\safesc{CompaSO}}
\newcommand{\abacushod}{\safesc{AbacusHOD}}

\newcommand{\supplementarylink}{\doi{10.5281/zenodo.16943122}}

\begin{document}

\title{Clustering of DESI galaxies split by thermal Sunyaev-Zeldovich effect}


\def\andname{}

\author{
M.~Rashkovetskyi\orcidlink{0000-0001-7144-2349}\altaffilmark{1,2,3,4,5},
D.~J.~Eisenstein\altaffilmark{1},
J.~Aguilar\altaffilmark{6},
S.~Ahlen\orcidlink{0000-0001-6098-7247}\altaffilmark{7},
A.~Anand\orcidlink{0000-0003-2923-1585}\altaffilmark{6},
D.~Bianchi\orcidlink{0000-0001-9712-0006}\altaffilmark{8,9},
D.~Brooks\altaffilmark{10},
F.~J.~Castander\orcidlink{0000-0001-7316-4573}\altaffilmark{11,12},
T.~Claybaugh\altaffilmark{6},
A.~Cuceu\orcidlink{0000-0002-2169-0595}\altaffilmark{6},
K.~S.~Dawson\orcidlink{0000-0002-0553-3805}\altaffilmark{13},
A.~de la Macorra\orcidlink{0000-0002-1769-1640}\altaffilmark{14},
Arjun~Dey\orcidlink{0000-0002-4928-4003}\altaffilmark{15},
P.~Doel\altaffilmark{10},
S.~Ferraro\orcidlink{0000-0003-4992-7854}\altaffilmark{6,16},
A.~Font-Ribera\orcidlink{0000-0002-3033-7312}\altaffilmark{17},
J.~E.~Forero-Romero\orcidlink{0000-0002-2890-3725}\altaffilmark{18,19},
E.~Gaztañaga\orcidlink{0000-0001-9632-0815}\altaffilmark{11,20,12},
G.~Gutierrez\altaffilmark{21},
H.~K.~Herrera-Alcantar\orcidlink{0000-0002-9136-9609}\altaffilmark{22,23},
K.~Honscheid\orcidlink{0000-0002-6550-2023}\altaffilmark{2,3,5},
C.~Howlett\orcidlink{0000-0002-1081-9410}\altaffilmark{24},
M.~Ishak\orcidlink{0000-0002-6024-466X}\altaffilmark{25},
R.~Joyce\orcidlink{0000-0003-0201-5241}\altaffilmark{15},
R.~Kehoe\altaffilmark{26},
T.~Kisner\orcidlink{0000-0003-3510-7134}\altaffilmark{6},
A.~Kremin\orcidlink{0000-0001-6356-7424}\altaffilmark{6},
O.~Lahav\altaffilmark{10},
A.~Lambert\altaffilmark{6},
M.~Landriau\orcidlink{0000-0003-1838-8528}\altaffilmark{6},
M.~Manera\orcidlink{0000-0003-4962-8934}\altaffilmark{27,17},
R.~Miquel\altaffilmark{28,17},
E.~Mueller\altaffilmark{29},
S.~Nadathur\orcidlink{0000-0001-9070-3102}\altaffilmark{20},
N.~Palanque-Delabrouille\orcidlink{0000-0003-3188-784X}\altaffilmark{23,6},
W.~J.~Percival\orcidlink{0000-0002-0644-5727}\altaffilmark{30,31,32},
F.~Prada\orcidlink{0000-0001-7145-8674}\altaffilmark{33},
I.~P\'erez-R\`afols\orcidlink{0000-0001-6979-0125}\altaffilmark{34},
A.~J.~Ross\orcidlink{0000-0002-7522-9083}\altaffilmark{2,4,5},
G.~Rossi\altaffilmark{35},
E.~Sanchez\orcidlink{0000-0002-9646-8198}\altaffilmark{36},
D.~Schlegel\altaffilmark{6},
M.~Schubnell\altaffilmark{37,38},
J.~Silber\orcidlink{0000-0002-3461-0320}\altaffilmark{6},
D.~Sprayberry\altaffilmark{15},
G.~Tarl\'{e}\orcidlink{0000-0003-1704-0781}\altaffilmark{38},
B.~A.~Weaver\altaffilmark{15},
R.~Zhou\orcidlink{0000-0001-5381-4372}\altaffilmark{6},
and H.~Zou\orcidlink{0000-0002-6684-3997}\altaffilmark{39}
}

\affil{$^{1}$ Center for Astrophysics $|$ Harvard \& Smithsonian, 60 Garden Street, Cambridge, MA 02138, USA}
\affil{$^{2}$ Center for Cosmology and AstroParticle Physics, The Ohio State University, 191 West Woodruff Avenue, Columbus, OH 43210, USA}
\affil{$^{3}$ Department of Physics, The Ohio State University, 191 West Woodruff Avenue, Columbus, OH 43210, USA}
\affil{$^{4}$ Department of Astronomy, The Ohio State University, 4055 McPherson Laboratory, 140 W 18th Avenue, Columbus, OH 43210, USA}
\affil{$^{5}$ The Ohio State University, Columbus, 43210 OH, USA}
\affil{$^{6}$ Lawrence Berkeley National Laboratory, 1 Cyclotron Road, Berkeley, CA 94720, USA}
\affil{$^{7}$ Department of Physics, Boston University, 590 Commonwealth Avenue, Boston, MA 02215 USA}
\affil{$^{8}$ Dipartimento di Fisica ``Aldo Pontremoli'', Universit\`a degli Studi di Milano, Via Celoria 16, I-20133 Milano, Italy}
\affil{$^{9}$ INAF-Osservatorio Astronomico di Brera, Via Brera 28, 20122 Milano, Italy}
\affil{$^{10}$ Department of Physics \& Astronomy, University College London, Gower Street, London, WC1E 6BT, UK}
\affil{$^{11}$ Institut d'Estudis Espacials de Catalunya (IEEC), c/ Esteve Terradas 1, Edifici RDIT, Campus PMT-UPC, 08860 Castelldefels, Spain}
\affil{$^{12}$ Institute of Space Sciences, ICE-CSIC, Campus UAB, Carrer de Can Magrans s/n, 08913 Bellaterra, Barcelona, Spain}
\affil{$^{13}$ Department of Physics and Astronomy, The University of Utah, 115 South 1400 East, Salt Lake City, UT 84112, USA}
\affil{$^{14}$ Instituto de F\'{\i}sica, Universidad Nacional Aut\'{o}noma de M\'{e}xico,  Circuito de la Investigaci\'{o}n Cient\'{\i}fica, Ciudad Universitaria, Cd. de M\'{e}xico  C.~P.~04510,  M\'{e}xico}
\affil{$^{15}$ NSF NOIRLab, 950 N. Cherry Ave., Tucson, AZ 85719, USA}
\affil{$^{16}$ University of California, Berkeley, 110 Sproul Hall \#5800 Berkeley, CA 94720, USA}
\affil{$^{17}$ Institut de F\'{i}sica d’Altes Energies (IFAE), The Barcelona Institute of Science and Technology, Edifici Cn, Campus UAB, 08193, Bellaterra (Barcelona), Spain}
\affil{$^{18}$ Departamento de F\'isica, Universidad de los Andes, Cra. 1 No. 18A-10, Edificio Ip, CP 111711, Bogot\'a, Colombia}
\affil{$^{19}$ Observatorio Astron\'omico, Universidad de los Andes, Cra. 1 No. 18A-10, Edificio H, CP 111711 Bogot\'a, Colombia}
\affil{$^{20}$ Institute of Cosmology and Gravitation, University of Portsmouth, Dennis Sciama Building, Portsmouth, PO1 3FX, UK}
\affil{$^{21}$ Fermi National Accelerator Laboratory, PO Box 500, Batavia, IL 60510, USA}
\affil{$^{22}$ Institut d'Astrophysique de Paris. 98 bis boulevard Arago. 75014 Paris, France}
\affil{$^{23}$ IRFU, CEA, Universit\'{e} Paris-Saclay, F-91191 Gif-sur-Yvette, France}
\affil{$^{24}$ School of Mathematics and Physics, University of Queensland, Brisbane, QLD 4072, Australia}
\affil{$^{25}$ Department of Physics, The University of Texas at Dallas, 800 W. Campbell Rd., Richardson, TX 75080, USA}
\affil{$^{26}$ Department of Physics, Southern Methodist University, 3215 Daniel Avenue, Dallas, TX 75275, USA}
\affil{$^{27}$ Departament de F\'{i}sica, Serra H\'{u}nter, Universitat Aut\`{o}noma de Barcelona, 08193 Bellaterra (Barcelona), Spain}
\affil{$^{28}$ Instituci\'{o} Catalana de Recerca i Estudis Avan\c{c}ats, Passeig de Llu\'{\i}s Companys, 23, 08010 Barcelona, Spain}
\affil{$^{29}$ Department of Physics and Astronomy, University of Sussex, Brighton BN1 9QH, U.K}
\affil{$^{30}$ Department of Physics and Astronomy, University of Waterloo, 200 University Ave W, Waterloo, ON N2L 3G1, Canada}
\affil{$^{31}$ Perimeter Institute for Theoretical Physics, 31 Caroline St. North, Waterloo, ON N2L 2Y5, Canada}
\affil{$^{32}$ Waterloo Centre for Astrophysics, University of Waterloo, 200 University Ave W, Waterloo, ON N2L 3G1, Canada}
\affil{$^{33}$ Instituto de Astrof\'{i}sica de Andaluc\'{i}a (CSIC), Glorieta de la Astronom\'{i}a, s/n, E-18008 Granada, Spain}
\affil{$^{34}$ Departament de F\'isica, EEBE, Universitat Polit\`ecnica de Catalunya, c/Eduard Maristany 10, 08930 Barcelona, Spain}
\affil{$^{35}$ Department of Physics and Astronomy, Sejong University, 209 Neungdong-ro, Gwangjin-gu, Seoul 05006, Republic of Korea}
\affil{$^{36}$ CIEMAT, Avenida Complutense 40, E-28040 Madrid, Spain}
\affil{$^{37}$ Department of Physics, University of Michigan, 450 Church Street, Ann Arbor, MI 48109, USA}
\affil{$^{38}$ University of Michigan, 500 S. State Street, Ann Arbor, MI 48109, USA}
\affil{$^{39}$ National Astronomical Observatories, Chinese Academy of Sciences, A20 Datun Road, Chaoyang District, Beijing, 100101, P.~R.~China}

\thanks{E-mail: \href{mailto:rashkovetskyi.1@osu.edu}{rashkovetskyi.1@osu.edu}}

\begin{abstract}
The thermal Sunyaev-Zeldovich (tSZ) effect is associated with galaxy clusters --- extremely large and dense structures tracing the dark matter with a higher bias than isolated galaxies.
We propose to use the tSZ data to separate galaxies from redshift surveys into distinct subpopulations corresponding to different densities and biases independently of the redshift survey systematics.
Leveraging the information from different environments, as in density-split and density-marked clustering, is known to tighten the constraints on cosmological parameters, like $\Omega_m$, $\sigma_8$ and neutrino mass.
We use data from the Dark Energy Spectroscopic Instrument (DESI) and the Atacama Cosmology Telescope (ACT) in their region of overlap to demonstrate informative tSZ splitting of Luminous Red Galaxies (LRGs).
We discover a significant increase in the large-scale clustering of DESI LRGs corresponding to detections starting from 1-2 sigma in the ACT DR6 + {\it Planck} tSZ Compton-$y$ map, below the cluster candidate threshold (4 sigma).
We also find that such galaxies have higher line-of-sight coordinate (and velocity) dispersions and a higher number of close neighbors than both the full sample and near-zero tSZ regions.
We produce simple simulations of tSZ maps that are intrinsically consistent with galaxy catalogs and do not include systematic effects, and find a similar pattern of large-scale clustering enhancement with tSZ effect significance.
Moreover, we observe that this relative bias pattern remains largely unchanged with variations in the galaxy-halo connection model in our simulations.
This is promising for future cosmological inference from tSZ-split clustering with semi-analytical models.
Thus, we demonstrate that valuable cosmological information is present in the lower signal-to-noise regions of the thermal Sunyaev-Zeldovich map, extending far beyond the individual cluster candidates.
\end{abstract}


\maketitle

\section{Introduction}

Modern cosmology is a high-precision science thanks to a rich variety of data collected.
The large-scale structure of the Universe contributes a major portion of this data.
Most important techniques include baryon acoustic oscillations distance measurements \citep[e.g.,][]{DESI.DR2.BAO.lya,DESI.DR2.BAO.cosmo} and full-shape clustering analyses \citep{DESI2024.V.KP5,DESI2024.VII.KP7B}, both based on galaxy redshift surveys.

Looking into the future, it is important to determine priorities for the next spectroscopic surveys.
The expansion towards higher redshifts is well motivated by the primordial Universe studies and supported by rigorous forecasts.
However, observations pose new challenges as spectral features of galaxies redshift out of the near-visible range accessible to ground-based telescopes \citep{snowmass2021-high-z-LSS}.

In contrast, measuring more galaxy spectra at lower redshifts is easier technically.
We already have a dense sample in the DESI Bright Galaxy Survey (\cite{BGS.TS.Hahn.2023}).
In the future DESI upgrade, DESI-II, the density of the luminous red galaxy (LRG) sample will also increase greatly \citep{spectroscopic-roadmap-cosmic-frontier}.
Expansion and growth history of the Universe at low and intermediate redshifts $z\lesssim 1$ is sensitive to the (late-time) dark energy \citep{DESI.DR2.BAO.cosmo} as well as curvature \citep{curvature-DESI-DR2-BAO} and even early dark energy \citep{EDE-DESI-DR2-BAO}.
However, smaller available volume at lower redshifts limits the large-scale information, motivating the push towards smaller, non-linear scales, with which robust results and forecasts have been very challenging.

There are promising ideas for leveraging information from different galaxy environments that can benefit from dense samples.
One is density-marked clustering, which assigns additional weight to galaxies based on a local density estimate at their location.
It was originally introduced in \cite{density-marked-CF-MG} to enhance modified gravity tests,
and was later tuned to give tighter constraints on neutrino mass \citep{density-marked-PS-neutrinos} and other cosmological parameters \citep{density-marked-PS-cosmoinfo,density-marked-PS-SBI}.
Analytical models of density-marked clustering were also proposed by \cite{density-marked-CFs-PT,density-marked-PS-PT,density-marked-PS-modeling,density-marked-PS-analytic}.
Another is density-split clustering, which splits galaxies into subsamples based on the local density estimate.
It was introduced for better modeling of redshift-space distortions in \cite{density-split-clustering-RSD}.
Later works predict tighter constraints on standard cosmological parameters and neutrino masses \citep{density-split-clustering-constrain-nuLCDM}, or primordial non-Gaussianity \citep{density-split-clustering-PNG}.
A simulation-based model of density-split clustering was built \citep{density-split-clustering-sim-based-model} and applied to BOSS CMASS data \citep{density-split-clustering-BOSS-CMASS}, surpassing the standard analysis.

A number of works demonstrated the benefit of combining galaxy surveys with data on secondary cosmic microwave background (CMB) anisotropies, particularly lensing.
For example, \cite{unWISE-galaxiesxACT-lensing,ACT-lensingxDESI-LRG-structure-formation-Kim,ACT-lensingxDESI-LRG-structure-formation-Sailer} cross-correlate lensing with photometric galaxies to better constrain the growth of cosmic structure, and \cite{DESI-QSOxPlanck-lensing-PNG,DESI-LRGxPlanck-lensing-PNG} improve measurements of primordial non-Gaussianity.
\cite{joint-clustering-lensing-Maus} present an even broader joint analysis of spectroscopic galaxy clustering and cross-correlation of photometric galaxies with CMB lensing to constrain the $z\sim 1$ clustering amplitude and test general relativity.
This combination of probes is very promising, because DESI is taking galaxy spectra at an unprecedented rate, and the next CMB experiments like Simons Observatory \citep{SO} and CMB-S4 \citep{CMBS4,CMBS4white} have great prospects for secondary CMB anisotropies, including lensing and Sunyaev-Zeldovich effects.

We decided to use the thermal Sunyaev-Zeldovich (tSZ) effect as a means of separating galaxies into different host environments.
The effect is produced by inverse Compton scattering of cosmic microwave background (CMB) photons on free thermal electrons moving randomly (whereas bulk motions give rise to the kinetic or kinematic Sunyaev-Zeldovich effect).
This process results in a net increase in energy of scattered photons and creates a distinct frequency-dependent distortion in the CMB spectrum.
Accordingly, CMB experiments observing in different frequency bands can disentangle tSZ emission from other components of the microwave sky.
The relative change in photon energy is approximately equal to the Compton $y$ parameter, which is proportional to the integral of free electron density $n_e$ and temperature $T_e$ along the path \citep[with length element $dr$;][]{Planck-SZ-map,Sunyaev-Zeldovich-1972}:
\begin{equation} \label{eq:y-parameter-0}
    y = \int \frac{k_B T_e}{m_e c^2} n_e \sigma_T dr,
\end{equation}
other quantities are constants: electron rest mass energy $m_e c^2$, Boltzmann's constant $T_e$ and Thomson scattering cross-section $\sigma_T$.
Accordingly, the effect is strongest in ionized, hot and dense gas in or around galaxy clusters \citep[as suggested by][]{Sunyaev-Zeldovich-1970,Sunyaev-Zeldovich-1980-review}.
The clusters represent a distinct environment.

The tSZ data comes from CMB telescopes independent of DESI, whereas density estimates depend on the galaxy observation strategies.
Using different datasets would allow us to glean complementary insights, check for systematics and potentially discover new fundamental tensions.
Specific improvements could include
testing the conformity of different galaxy sub-types \citep[e.g.,][]{LSS-color-dependent-stochasticity},
improving the redshift-space distortion modeling for full-shape clustering analyses by removing a small fraction of the strongest Fingers of God \citep{removing-FoG},
obtaining multiple tracers with better-constrained biases for the primordial non-Gaussianity measurements \citep[similarly to][]{multi-tracer-PNG-forecasts}
or constraining environmental effects on galaxy formation or galaxy-halo connection \citep[e.g.,][]{EDR_HOD_LRGQSO2023}.

We need to remark that Sunyaev-Zeldovich maps have been used for cluster studies: their detection, mass determination, and more \citep{Planck-SZ-clusters,ACT-SZ-clusters-DR5,ACT-SZ-clusters-mass-calibration-DR5,SPT-clusters-w-DES+HST-WL,SPT-SZ-clusters-2025,ACT-SZ-clusters-DR6}.
But rigorously detected and confirmed cluster candidates are rare.
We aim to extract more information from the lower signal-to-noise parts, which comprise a much bigger fraction of the map.

This work is not going to provide the final answers and methodology, but rather motivate further developments.
It is structured as follows:
\cref{sec:data} introduces the data and simulations we use,
\cref{sec:measurements} details our processing of real data and provides its results,
\cref{sec:mock} describes our simulation-based toy model, compares it with data and shows new insights,
\cref{sec:conclusions} concludes with a summary and a future outlook.

\section{Data}
\label{sec:data}

\subsection{Galaxy catalog: DESI DR1 LRG}

The Dark Energy Spectroscopic Instrument \citep[DESI][]{DESI2016b.Instr,DESI2022.KP1.Instr} is a robotic fiber spectrograph operating on the 4-meter Mayall telescope at Kitt Peak National Observatory.
Since 2021, it conducts an 8-year galaxy survey of approximately 40\% of the sky, which will yield nearly 63 million galaxy and quasar redshifts \citep[compared to the original forecast of 39 million for the 5-year program in][]{Snowmass2013.Levi,DESI2016a.Science}.
This efficiency is enabled by instrumentation and software innovations, including a focal plane design with the robotic positioners for approximately 5000 optical fibers \citep{FocalPlane.Silber.2023,FiberSystem.Poppett.2024}, an optical corrector widening the field of view to $\sim 3^\circ$, a comprehensive spectroscopic reduction pipeline \citep{Spectro.Pipeline.Guy.2023} and a pipeline to adjust observation planning and optimization as the survey progresses \citep{SurveyOps.Schlafly.2023}.

DESI's scientific program has been successfully validated \citep{DESI2023a.KP1.SV} alongside the Early Data Release \citep{DESI2023b.KP1.EDR}.
The key results based on Data Release 1 \citep{DESI2024.I.DR1} include clustering catalogs and two-point statistics measurements \citep{DESI2024.II.KP3}, baryon acoustic oscillation (BAO) distance measurements from galaxies, quasars \citep{DESI2024.III.KP4} and Lyman-$\alpha$ forest \citep{DESI2024.IV.KP6} along with their detailed implications for cosmological models \citep{DESI2024.VI.KP7A}; and full-shape clustering of galaxies and quasars \citep{DESI2024.V.KP5} with their cosmological analysis \citep{DESI2024.VII.KP7B}.
Furthermore, updated BAO measurements from Lyman-$\alpha$ \citep{DESI.DR2.BAO.lya}, galaxies and quasars with accompanying cosmological interpretations \citep{DESI.DR2.BAO.cosmo} are also available, based on Data Release 2 \citep{DESI.DR2.DR2}.

We use Luminous Red Galaxies \citep[LRG;][]{LRG.TS.Zhou.2023} from the DESI DR1 clustering catalog covering 5,740~deg$^2$ \citep{DESI2024.II.KP3}.
We chose LRG because their redshift range ($0.4<z<1.1$) covers the peak in redshift distribution of Sunyaev-Zeldovich clusters \citep{ACT-SZ-clusters-DR6}.
We discard galaxies at $z>0.85$ because then both the LRG density \citep{DESI2024.II.KP3} and the SZ cluster abundance \citep{ACT-SZ-clusters-DR6} drop.

\subsection{Sunyaev-Zeldovich map: ACT DR6}

The Atacama Cosmology Telescope \citep[ACT;][]{ACT-design,ACT} was a ground-based cosmic microwave background (CMB) experiment.
Compared to the space-based {\it Planck} mission \citep{planck_overview}, it has higher resolution and lower instrumental noise, but a smaller sky fraction and fewer frequency bands (which makes component separation harder).
ACT's advantages enable smaller-scale Sunyaev-Zeldovich measurements, which can be combined with robust larger-scale results from {\it Planck}.

We use the thermal Sunyaev-Zeldovich Compton $y$ parameter map from ACT Data Release 6 covering $\approx 13{,}000$~deg$^2$ \citep{ACT-component-separated-maps-DR6}, which builds upon {\it Planck} data \citep{Planck-SZ-map}.
We rely on the accompanying noise simulations \citep{ACT-noise-simulations-DR6} to estimate the signal-to-noise ratio in our analysis.
We primarily work in the $\approx 3{,}510$~deg$^2$ area of overlap between DESI DR1 LRG and ACT DR6 footprints.

\subsection{Simulations: \abacussummit{} halo catalogs and \abacushod{} galaxy-halo connection model}

\defcitealias{Zheng2007-HOD}{Zheng07}

We also use a $z=0.8$ snapshot of a $\qty(2\ihGpc)^3$ cubic box (for the fiducial cosmology) from the \abacussummit{} suite of $N$-body simulations \citep{AbacusSummit} produced with \abacus{} code \citep{Abacus-code}.
The halos have been identified with the \compaso{} halo finder \citep{CompaSO-halo-finder}.
We use a galaxy catalog produced within the halo occupation distribution (HOD) galaxy-halo connection framework, efficiently implemented in \abacushod{} \citep{AbacusHOD} with parameters based on \cite{EDR_HOD_LRGQSO2023} (\citetalias{Zheng2007-HOD}$+f_{\rm ic}+\alpha_{\rm c}+\alpha_{\rm s}$ model, but with $f_{\rm ic}=1$).
This galaxy catalog was one of the base cubic boxes for {\tt Abacus-2} cut-sky mocks described in \cite{DESI2024.III.KP4}.

\section{Measurements}
\label{sec:measurements}

\subsection{Methodology and challenges}
\label{sec:measurements:methodology}

We divide the LRGs into subsamples (``SNR bins'') according to the signal-to-noise ratio at their location in the ACT DR6 + {\it Planck} Compton-$y$ map \citep{ACT-component-separated-maps-DR6}.
First, we match the positions of the LRGs from the DESI clustering catalog to the pixels in the map\footnote{The ACT DR6 map is provided as a rectangular grid in DEC and RA with a pixel side of 0.5 arcmin.}.
We only use the pixels with the apodized mask value above the threshold of 0.9 (this slightly reduces the footprint from $\approx 3{,}510$~deg$^2$ to $\approx 3{,}420$~deg$^2$).
Second, we compute the pixel-level standard deviations in $y$ using the corresponding 304 Gaussian noise simulation \citep{ACT-noise-simulations-DR6}.
Finally, we compute the signal-to-noise ratio in each pixel by dividing the $y$ value by its standard deviation.

Such external selection of galaxies imposes complex geometry changes that affect the estimation of the clustering statistics.
E.g., the Landy-Szalay estimator \citep{Landy-Szalay} for the correlation function between samples 1 and 2 (which may be identical or different):
\begin{equation} \label{eq:2PCF-Landy-Szalay}
    \hat \xi^{\rm LS}_{12} = \frac{D_1 D_2 - D_1 R_2 - R_1 D_2}{R_1 R_2} + 1,
\end{equation}
where $D_1 D_2$ are the (binned) pair counts between data (galaxies) from sample 1 and data 2, $D_1 R_2$ are the pair counts between data 1 and random points (reflecting the survey geometry and selection etc) for sample 2, $R_1 D_2$ are between random points 1 and data 2, and $R_1 R_2$ are between random points 1 and 2.

The key to the problem is that the full-survey random catalogs are not representative of our subsamples.
One might think that imposing the same on-sky position filter on random points would be a solution; however, it is not perfect.
The resulting randoms would live in (narrow) ``cylinders'' extended along the line of sight, whereas a significant portion of selected galaxies would be truly clumped around the cluster position in a three-dimensional sense.

It is possible to accept the unusual-looking clustering measurement and model it consistently, but we have decided to keep the correlation functions more intuitive.
To achieve this, we avoid the issue with the subsample geometry by employing an asymmetric Davis-Peebles estimator \citep{Davis-Peebles} for the correlation function:
\begin{equation} \label{eq:2PCF-Davis-Peebles}
    \hat \xi^{\rm DP}_{12} = \frac{D_1 D_2}{D_1 R_2} - 1.
\end{equation}
The $D_1 D_2$ and $D_1 R_2$ are the same pair counts as in the Landy-Szalay estimator (\cref{eq:2PCF-Landy-Szalay}).
However, we note that the randoms representing sample 1 are not required: neither $R_1 D_2$ nor $R_1 R_2$ are involved.
The randoms are only necessary for sample 2.
Therefore, with this estimator, we can compute the correlation function between the tSZ subsample (as 1) and the full LRG sample (as 2).

We also apply filters to the Compton-$y$ map for two reasons.
First, a filter matched to a cluster SZ profile would optimally detect them and recover their positions \citep{matched-filters-intro}.
Second, the environmental influence of a cluster is likely to extend further than its Compton parameter profile.
Our fiducial filter is not thoroughly matched, but has a Gaussian shape with a 2.4 arcmin full width at the half-maximum (corresponding to $\approx 1.1~\ihMpc$ at $z=0.6$) for simplicity.
The single reference filter in the ACT DR6 SZ cluster paper has the same scale \citep{ACT-SZ-clusters-DR6}.
We apply the filter by transforming the map to spherical harmonics $a_{lm}$, multiplying them by the Gaussian filter profile $f_l$, and transforming back to the sky positions.
We apply the same filter to all the noise simulations to obtain the corresponding pixel-wise standard deviation map.
For reference, we provide galaxy counts in each signal-to-noise ratio bin after filtering in \cref{tab:galaxy-numbers-SNR-bins}.

\begin{table}[tbp]
    \centering
    \begin{tabular}{|r|r|}
        \hline
        SNR bin & $N_{\rm LRG}$ \\
        \hline
        $(-\infty, -2)$ $\sigma$ & 25101 \\
        $[-2, -1)$ $\sigma$ & 130331 \\
        $[-1, 0)$ $\sigma$ & 308498 \\
        $[0, 1)$ $\sigma$ & 305819 \\
        $[1, 2)$ $\sigma$ & 127875 \\
        $[2, 3)$ $\sigma$ & 25469 \\
        $[3, 4)$ $\sigma$ & 3577 \\
        $[4, 5)$ $\sigma$ & 981 \\
        $[5, 6)$ $\sigma$ & 484 \\
        $[6, \infty)$ $\sigma$ & 676 \\
        \hline
        Total & 928811 \\
        \hline
    \end{tabular}
    \caption{Number of DESI DR1 LRGs ($0.4<z<0.85$ and in the overlap with ACT footprint) in different tSZ SNR bins (after applying our fiducial Gaussian filter to the Compton-$y$ map).}
    \label{tab:galaxy-numbers-SNR-bins}
\end{table}

We should remark on the large areas with negative Compton $y$ parameter, contradicting its theoretical definition (\cref{eq:y-parameter-0}).
Instrumental (or atmospheric) noise plays a certain part.
Kinematic Sunyaev-Zeldovich effect, resulting from the bulk motions of free electrons, can have different signs and can become partially confused with ``negative'' thermal effect.
It is important to note that these factors produce chance fluctuations and not systematic biases, provided that the instrument bandpasses are modeled accurately.
Other components, like the cosmic infrared background (CIB), may also leak into the thermal Sunyaev-Zeldovich reconstruction, potentially biasing the tSZ measurements \citep[e.g.,][]{stacked-tSZ-profiles-ACT-DESI-LRG}.
CIB-deprojected maps are available for ACT DR6 \citep{ACT-component-separated-maps-DR6}, but in multiple variants with different parameters, so we defer the consistency check with them to future work.

\begin{figure}[btp]
    \centering
    \includegraphics[width=\linewidth]{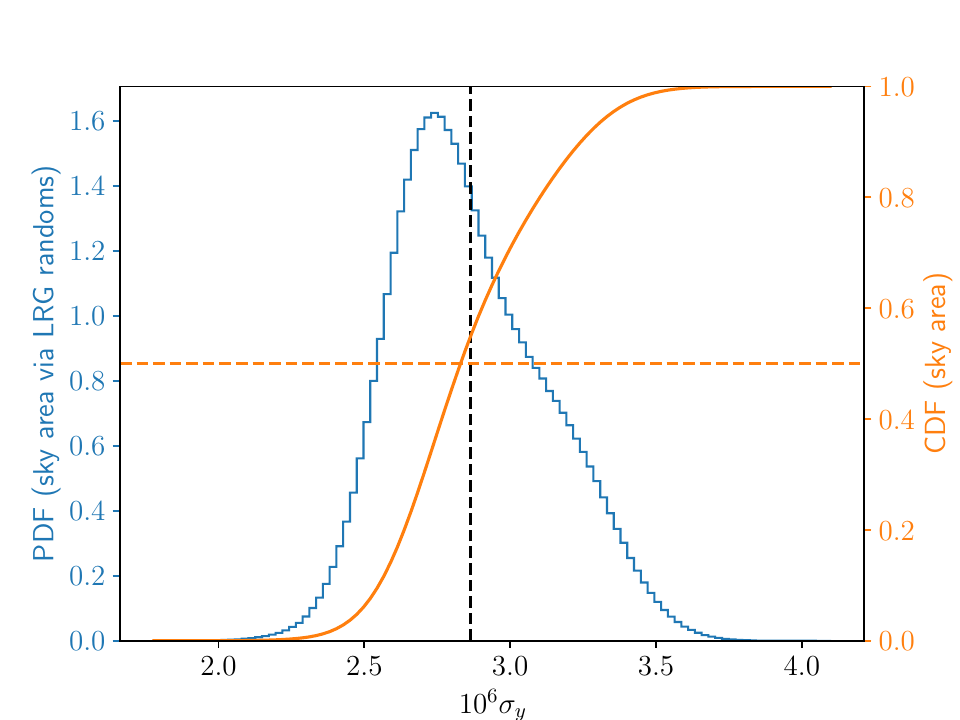}
    \caption{Distribution of $\sigma_y$, the pixel standard deviation of Compton-$y$ parameter after our filtering (over the area of the overlap of ACT DR6 and DESI DR1 footprints).
    The black dashed vertical line shows the mean $\sigma_y$.
    The orange dashed horizontal line denotes CDF$=0.5$.}
    \label{fig:y-std-dist}
\end{figure}

Finally, we briefly comment on our choice of tSZ signal-to-noise ratio ($y/\sigma_y$) instead of the signal (Compton-$y$).
The signal can be more directly related to the mass of the cluster or the dark matter halo in which the galaxy resides\footnote{But note that our simple filter does not exactly match the filters used in tSZ cluster studies like \cite{ACT-SZ-clusters-DR5,ACT-SZ-clusters-mass-calibration-DR5,SPT-SZ-clusters-2025,ACT-SZ-clusters-DR6}, so the relations calibrated there do not directly apply to our measurements.}.
However, we want to study the low signal-to-noise regime, in which a measured Compton-$y$ inevitably corresponds to a relatively wide range of true Compton-$y$ parameters.
Signal-to-noise ratio better characterizes the fraction of near-zero signal objects shifted up (or down) by noise, and we find it a more significant advantage.
Moreover, the range of pixel-wise standard deviations of the Compton-$y$ parameter within the footprint (shown in \cref{fig:y-std-dist}) is quite narrow.
Accordingly, we find very little difference between comparable bins in SNR ($y/\sigma_y$) and signal (Compton-$y$).

\subsection{Larger-scale clustering and galaxy bias}
\label{sec:measurements:large}

First, we compute the larger-scale cross-correlation functions between the different SNR bins and the full LRG sample.
We use the radial ($s$) and angular ($\mu$) binning.
We disregard the pairs located too closely on the sky\footnote{This is technically known as the theta ($\theta$) cut. Note that here $\theta$ is the angle between the lines of sight from the Solar System to the two galaxies, and not the angle $\Theta$ between the line of sight and the vector from one galaxy to the other used for the binning of the correlation function in $\mu \equiv \cos\Theta$.} for three reasons.
The first is to focus on the larger-scale, 2-halo clustering (we study the 1-halo regime in \cref{sec:measurements:small-LoS}).
The second is to avoid using pairs of galaxies belonging to the same SZ pixel, or associated with the same line of sight smeared by the beam and the filter.
This may also reduce sensitivity to cosmic infrared background contamination \citep[found significant in cluster profiles at smaller radii in][]{stacked-tSZ-profiles-ACT-DESI-LRG}.
The third is to mitigate the DESI fiber assignment incompleteness effects \citep{DESI2024.II.KP3,KP3s6-Bianchi}, for which \cite{KP3s5-Pinon} recommended the lower threshold of 0.05 degrees.
We only count pairs with the angular separation above 0.1 degrees (6 arcmin)\footnote{This leaves $\mu \approx 1$ bins with no pair counts and undefined correlation functions. We discard these bins and average over the remaining bins to obtain the correlation function monopole.}, which corresponds to the perpendicular separation $r_p\approx 2.7~\ihMpc$ at $z=0.6$\footnote{Ranging between $r_p\approx 1.9~\ihMpc$ at $z=0.4$ and $r_p\approx 3.5~\ihMpc$ at $z=0.85$}.

\begin{figure}[tbp]
    \centering
    \includegraphics[width=\linewidth]{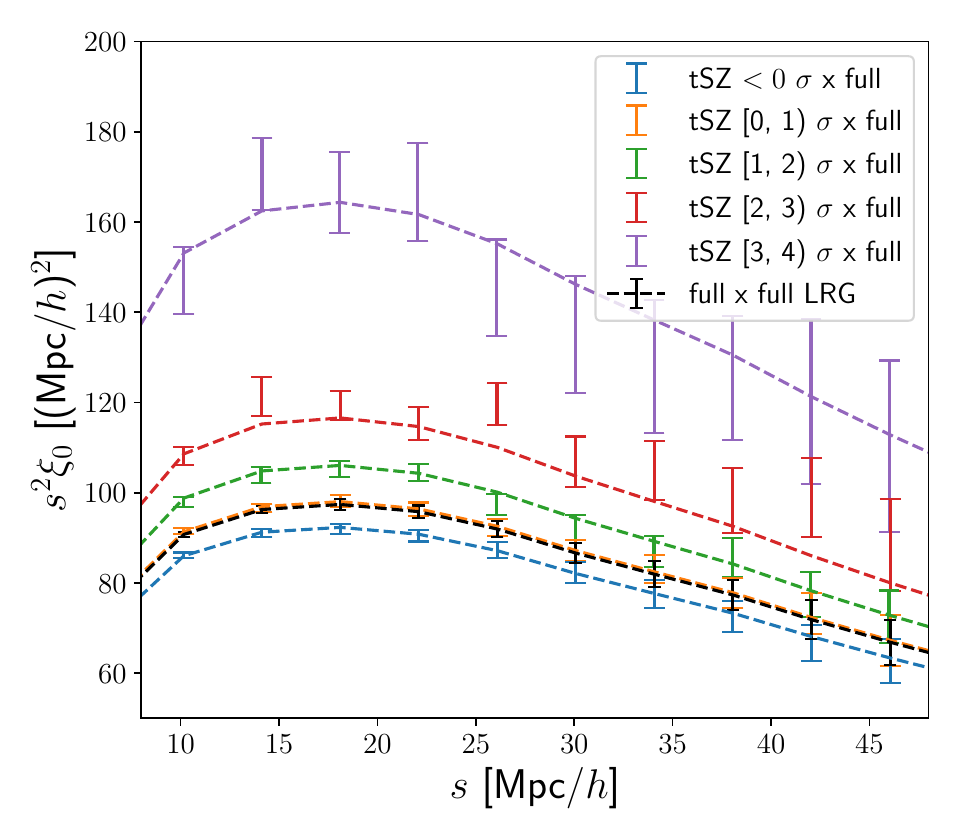}
    \caption{Large-scale isotropic (monopole) cross-correlation functions of different SNR bins with the full LRG sample (excluding pairs with angular separation below 0.1 degrees).
    (In this and the following figures, we apply our fiducial Gaussian filter to the tSZ Compton-$y$ parameter map, and use the jackknife technique to estimate errorbars.)
    There is a significant clustering enhancement with increasing tSZ detection level even below the threshold for cluster candidates \citep[$4\sigma$ in][]{ACT-SZ-clusters-DR6}.
    Colored dashed lines show the best fits obtained by scaling the full-sample autocorrelation function (black dashed line).
    The scaling coefficients are the relative biases shown in \cref{fig:relbias-SNR-bins}.}
    \label{fig:clustering-SNR-bins}
\end{figure}

We show the resulting isotropic cross-correlation functions in \cref{fig:clustering-SNR-bins} with covariances estimated using the jackknife technique.
We see a significant clustering enhancement in the monopole as the tSZ detection level increases, even though we stop at $4\sigma$, which was the threshold for cluster candidates in \cite{ACT-SZ-clusters-DR6}\footnote{However, bear in mind the simplicity of our filter, the matched filters of \cite{ACT-SZ-clusters-DR6} should be more optimal.}.
This almost certainly corresponds to an increase in galaxy bias.

The 0 to 1 $\sigma$ subsample appears very similar to the full LRG sample.
This is not explained by the vast majority of galaxies belonging to that bin, as \ref{tab:galaxy-numbers-SNR-bins} shows it contains just $\approx 1/3$ of our LRGs.
Instead, the galaxies with negative signal and higher SNR compensate each other so that their environmental characteristics average close to this $[0,1)$ $\sigma$ bin.
Interestingly, this similarity holds in all other aspects we consider in this paper.

For clarity, we seek to summarize each line in \cref{fig:clustering-SNR-bins} with a single number.
A simple way is to find the scaling for the full LRG correlation function to best match the cross-correlation function with any given signal-to-noise bin.
This multiplier should be similar to the ratio of the linear galaxy biases between the SNR subsample and the full LRG sample.
Accordingly, we also refer to this scaling as relative galaxy bias.

The jackknife covariance estimates for each correlation function allow us to estimate the precision of the scaling.
However, we have not estimated the covariances between different cross-correlation functions shown in \cref{fig:clustering-SNR-bins} consistently.
This makes our relative bias errorbar estimates imperfect and approximate.

\begin{figure}[tbp]
    \centering
    \includegraphics[width=\linewidth]{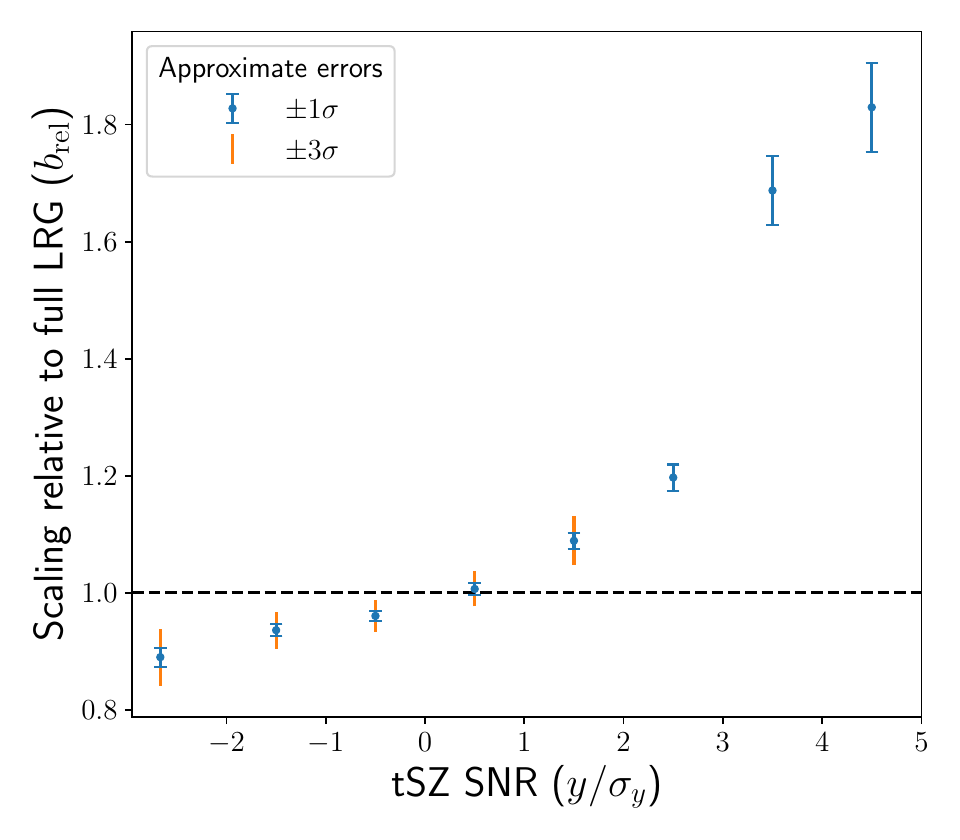}
    \caption{Galaxy bias of different SNR bins relative to the full LRG sample.
    These numbers are based on the ratio of the corresponding correlation functions from \cref{fig:clustering-SNR-bins}.
    The bias increase (clustering enhancement) can be seen more concisely here.
    For smaller errorbars, which can not be seen well, we also show $3\sigma$ bars (exactly 3 times larger).}
    \label{fig:relbias-SNR-bins}
\end{figure}

We show the resulting galaxy biases of our subsamples (relative to the full LRG sample) in \cref{fig:relbias-SNR-bins}.
The picture is consistent with the conclusions we have drawn from \cref{fig:clustering-SNR-bins}: the increase of galaxy bias with the tSZ detection level, and the 0 to 1 $\sigma$ bin being close to average over all LRGs.
In addition, we can inspect more categories at negative signal-to-noise, different from the full LRG but not so significantly deviating from each other.

\begin{figure}[tbp]
    \centering
    \includegraphics[width=\linewidth]{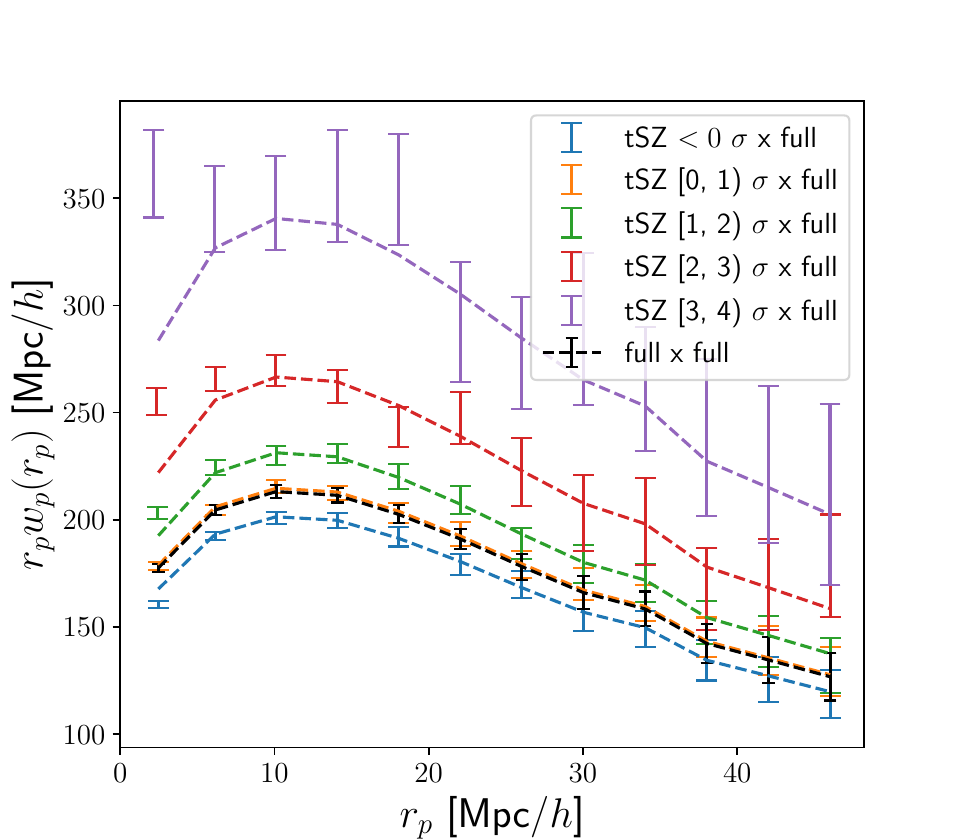}
    \caption{Projected cross-correlation functions of different SNR bins with the full LRG sample.
    The line-of-sight separation limit is $\pi_{\max}=50~\ihMpc$ (and we do not exclude pairs with small angular separations).
    Colored dashed lines show the best fits (using $r_p > 8~\ihMpc$) obtained by scaling the full-sample autocorrelation function (black dashed line).
    The clustering enhancement is notable, as with the correlation function monopole (\cref{fig:clustering-SNR-bins}).
    The relative biases obtained from projected clustering are consistent with \cref{fig:relbias-SNR-bins}, but have larger errorbars.}
    \label{fig:projected-clustering-SNR-bins}
\end{figure}

We also show the projected cross-correlation functions $w_p(r_p)$ in \cref{fig:projected-clustering-SNR-bins}.
We have computed it with the line-of-sight separation upper limit $\pi_{\max}=50~\ihMpc$ and without excluding pairs with small angular separation.
The cross-correlation functions still resemble the scaled full-sample auto-correlation on larger scales ($r_p \gtrsim 8~\ihMpc$).
The resulting relative bias values are close to our results from the 3D monopoles (\cref{fig:relbias-SNR-bins}).
This means that the pattern of stronger clustering with higher tSZ SNR is consistent between the two statistics.

The projected correlation function is beneficial because it marginalizes over redshift-space distortions, simplifying the modeling and interpretation.
On the flip side, we obtain slightly larger errorbars on relative bias obtained from projected clustering, but this depends on the ranges, which we have not optimized thoroughly in this proof-of-concept work.
In addition, precise redshifts are not critical for projected statistics.
This can allow future usage of a larger ``extended'' photometric LRG sample from DESI Legacy Surveys imaging \citep{LRG-samples-for-cross-correlations}, which also has more overlap area with ACT.

\subsection{Small-scale line-of-sight clustering and velocity dispersions}
\label{sec:measurements:small-LoS}

The cluster (supercluster) peculiar velocities are most apparent in configuration space at small scales near the line of sight.
Here we switch the correlation function bins from $s$ and $\mu$ to the line-of-sight ($\pi$) and the perpendicular (on-sky, $r_p$) separations between the galaxy pair members.
In this section, we also do not discard pairs with small angular separations, so the resulting clustering measurements are likely affected by DESI fiber assignment incompleteness \citep{DESI2024.II.KP3,KP3s6-Bianchi,KP3s5-Pinon}.

\begin{figure}[btp]
    \centering
    \includegraphics[width=\linewidth]{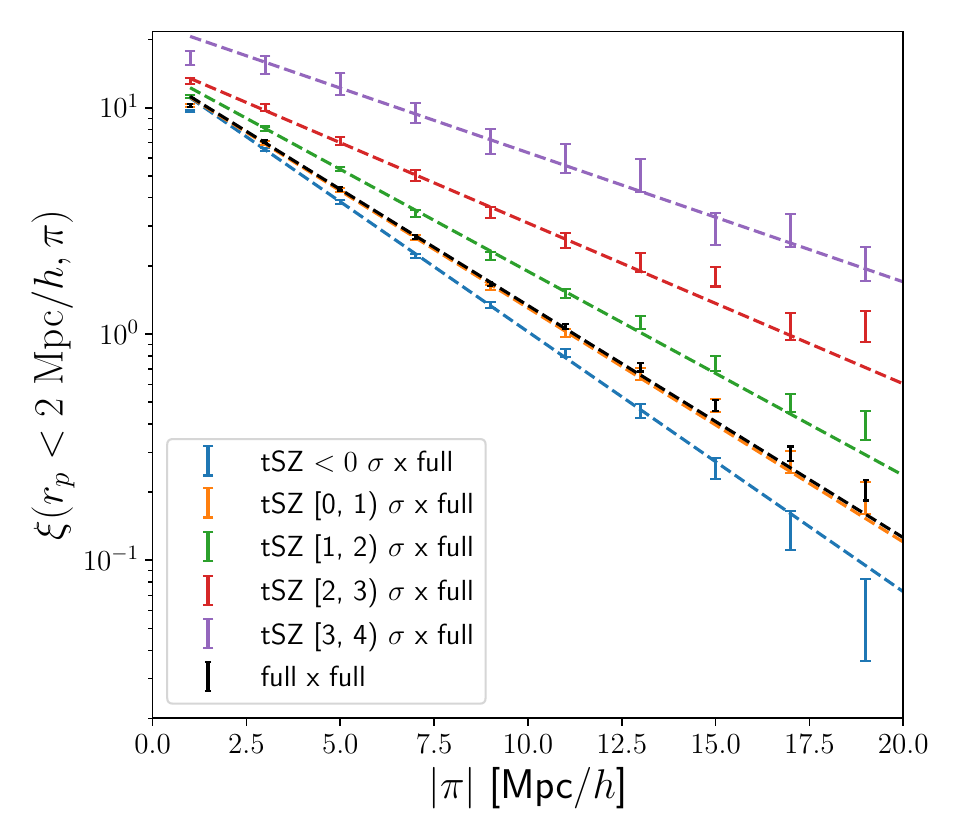}
    \caption{Small-scale line-of-sight cross-correlation functions of different SNR bins with the full LRG sample (not excluding pairs with small angular separations).
    The dashed lines show the best-fit exponentials according to \cref{eq:xi-LoS-exponential}.
    There is not only an increase in amplitude, but also a flattening of the slopes of the curves with increasing tSZ SNR, indicating regions with higher tSZ signal have hotter small-scale velocity dispersions.}
    \label{fig:LoS-clustering-SNR-bins}
\end{figure}

Accordingly, we plot the line-of-sight small-scale correlation functions for our SNR bins in \cref{fig:LoS-clustering-SNR-bins}.
We notice an increase in correlation functions with the tSZ detection level in each bin, similar to the trends in larger-scale correlation functions (\cref{fig:clustering-SNR-bins}).
More interestingly, the slopes of the curves are also noticeably different, although it is hard to judge the significance of these differences visually.

To better quantify the slopes, we fit exponentials to the correlation functions:
\begin{equation}
    \xi(\pi) \approx C \exp(-\sqrt{2} \frac{\abs{\pi}}{\sigma_\pi}). \label{eq:xi-LoS-exponential}
\end{equation}
$\sigma_\pi$ gives the one-dimensional dispersion (standard deviation).
This is motivated by the exponential fit to the distributions of peculiar radial velocities \citep{Peebles-cosmic-virial-theorem} and a physical Press--Schechter (Gaussian mixture) model \citep{pairwise-peculiar-velocity-distribution-nonlinear}.
The peculiar radial velocity $v_\parallel$ is related to the line-of-sight displacement between real and redshift space $\Delta \pi$ via \citep{Kaiser-RSD,effective-theories-peculiar-velocities}
\begin{equation}
    \Delta\pi = \frac{1+z}{H\qty(z)} v_\parallel.
\end{equation}
Because we have a wide redshift range ($z=0.4-0.85$), we can not use this equation for exact conversion.
However, we obtain approximate velocity dispersion figures by substituting the middle redshift $z_{\rm mid}=0.625$:
\begin{equation}
    \sigma_{1d} \approx \frac{H\qty(z_{\rm mid})}{1+z_{\rm mid}} \sigma_\pi. \label{eq:velocity-dispersion-approx}
\end{equation}

\begin{figure}[btp]
    \centering
    \includegraphics[width=\linewidth]{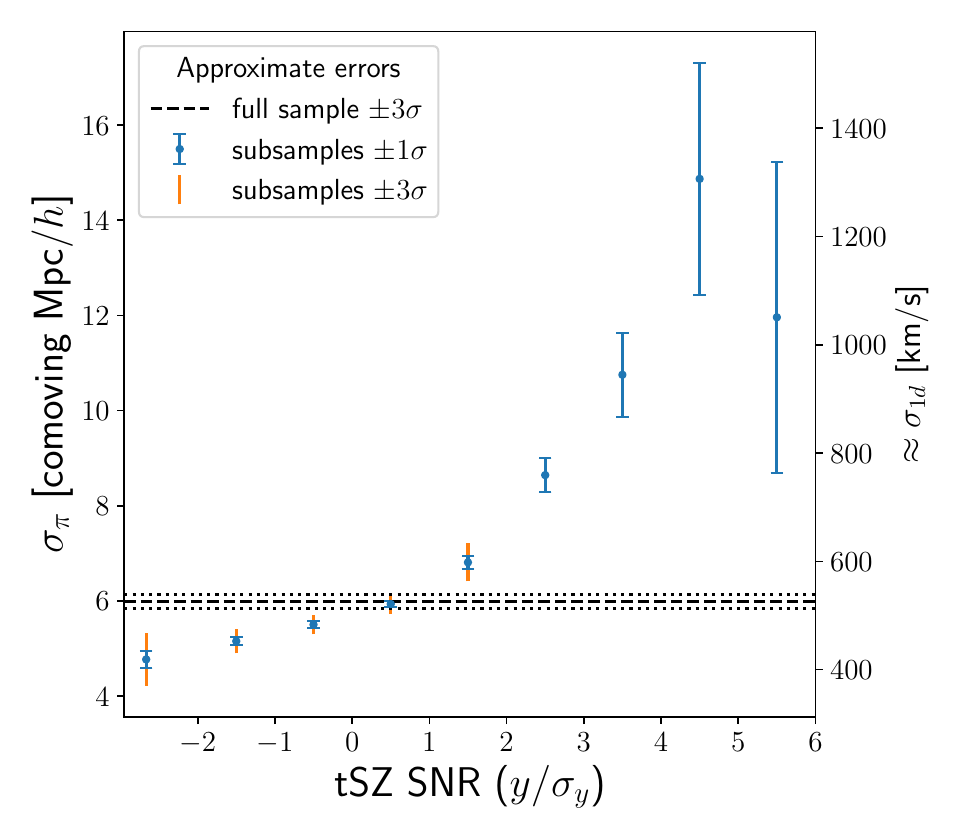}
    \caption{One-dimensional comoving coordinate dispersions $\sigma_\pi$ from small-scale line-of-sight clustering (\cref{fig:LoS-clustering-SNR-bins}) via \cref{eq:xi-LoS-exponential}.
    Approximate conversion to one-dimensional (line-of-sight) velocity dispersions $\sigma_{1d}$ via \cref{eq:velocity-dispersion-approx} is provided on the right.
    The dispersion values increase with the signal-to-noise ratio, except the 5 to 6 $\sigma_y$ bin.
    For smaller errorbars, which can not be seen well, we also show $3\sigma$ bars (exactly 3 times larger).}
    \label{fig:LoS-sigmas-SNR-bins}
\end{figure}

We plot the results of fitting \cref{eq:xi-LoS-exponential} for each tSZ SNR bin in \cref{fig:LoS-sigmas-SNR-bins}.
We measure them in comoving line-of-sight separations, but also provide an approximate conversion to line-of-sight peculiar velocities using the average redshift.
The dispersions increase with the tSZ detection level, more significantly for positive values\footnote{The last, 5 to 6 $\sigma_y$ bin, is an exception, although it has a large errorbar}.

The increases in velocity dispersion (\cref{fig:LoS-sigmas-SNR-bins}) and galaxy bias (\cref{fig:relbias-SNR-bins}) should be related.
Assuming the relation between halo mass and tSZ observables, they are two differently weighted integrals over the halo mass function.
Consistency of these patterns could be a powerful test of the $M_{\rm halo}-y$ relation, or systematics, but the DESI fiber assignment \citep{KP3s6-Bianchi} brings further complications.

The 0 to 1 $\sigma_y$ bin is again very close to the full LRG sample.
This is interesting because the full sample presumably includes very large clusters with extremely strong Fingers of God, unlike the $[0, 1)$ SNR bin.
However, the large cluster contribution should be suppressed by the DESI fiber assignment incompleteness \citep{KP3s6-Bianchi}, which could partially explain the similarity.

\begin{figure}[btp]
    \centering
    \includegraphics[width=\linewidth]{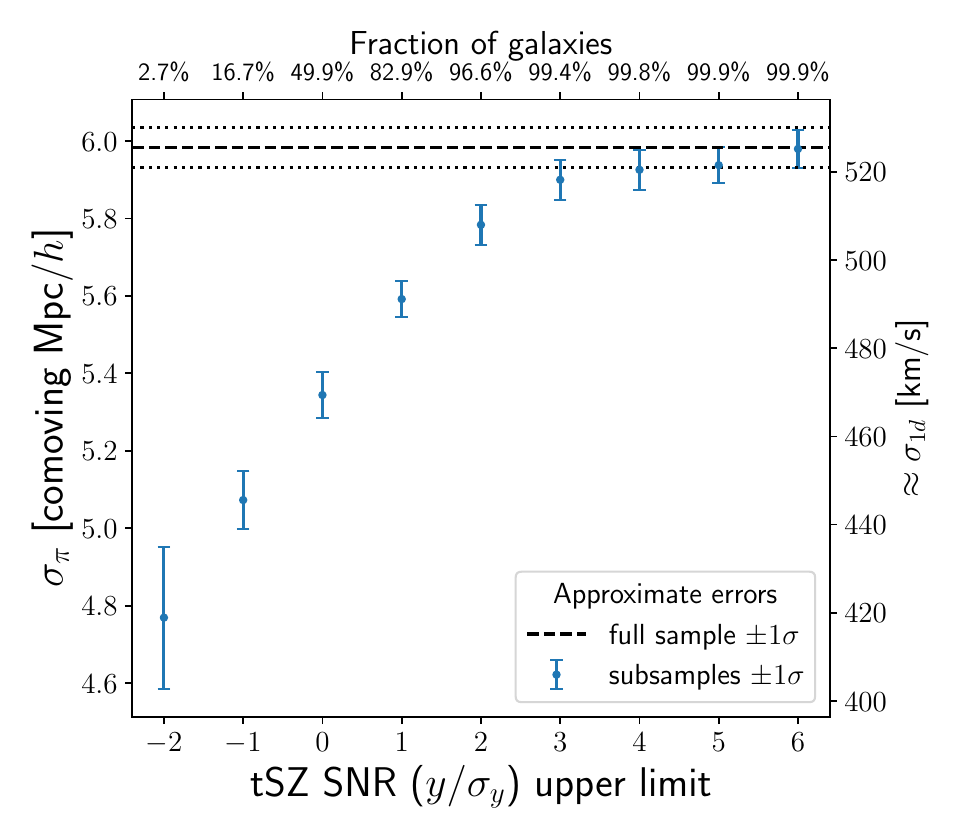}
    \caption{One-dimensional comoving coordinate dispersions $\sigma_\pi$ from small-scale line-of-sight clustering for subsamples strictly below a given tSZ SNR (``cumulative bins'') obtained by fitting \cref{eq:xi-LoS-exponential}.
    Approximate conversion to one-dimensional (line-of-sight) velocity dispersions $\sigma_{1d}$ via \cref{eq:velocity-dispersion-approx} is provided on the right.
    On top, we list fractions of the full sample belonging to each subsample.}
    \label{fig:LoS-sigmas-cumulative-SNR-bins}
\end{figure}

Following \cite{removing-FoG}, we also consider the lessening of Fingers of God by putting an upper limit on galaxies' tSZ SNR.
We construct subsamples strictly below several tSZ SNR values (``cumulative bins''), compute their line-of-sight clustering and extract coordinate dispersions $\sigma_\pi$.
We show the results in \cref{fig:LoS-sigmas-cumulative-SNR-bins}.
It shows that the velocity dispersion could be lowered by $\approx 20\%$ (at tSZ SNR $<-2$), but at the cost of discarding $\approx 97\%$ of galaxies.
The optimal cutoff is probably closer to an SNR of 1 or 2.

\begin{figure}[btp]
    \centering
    \includegraphics[width=\linewidth]{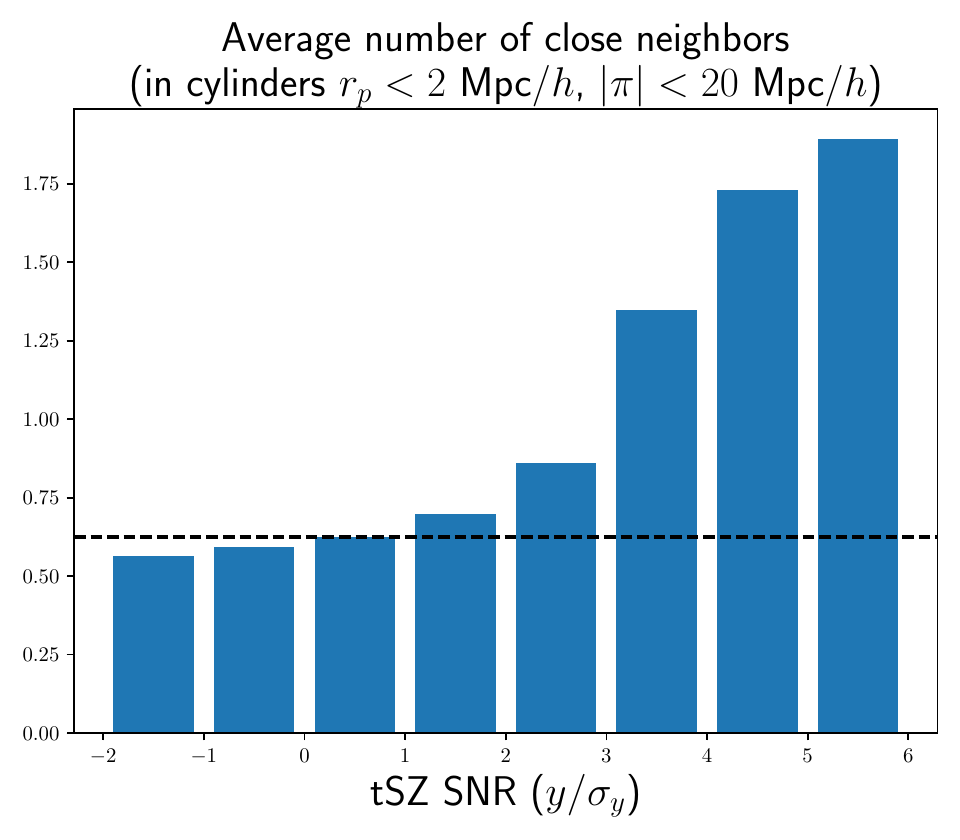}
    \caption{Average number of close neighbors in different tSZ SNR bins.
    (No lower limit on angular separation is applied.)
    It increases, confirming that stronger tSZ signatures are associated with clusters that have more galaxies.
    The dashed horizontal line shows the full-sample average number of neighbors, which is again very close to the $[0,1)$ $\sigma$ subsample.}
    \label{fig:avg-no-close-neighbors-cylinders-SNR-bins}
\end{figure}

Another natural way to summarize \cref{fig:LoS-clustering-SNR-bins} would be by integrating the correlation functions.
This is directly related to the average number of any LRGs in narrow cylinders with axes along the line of sight centered on LRGs belonging to a particular SNR bin.
We provide this summary in \cref{fig:avg-no-close-neighbors-cylinders-SNR-bins}.
The average number of neighbors increases, in accordance with the expectation that LRGs with higher tSZ SNR tend to live in larger clusters.

We have found even more striking differences in the abundance of galaxies with many neighbors in different SNR bins.
However, it is hard to visualize and interpret more meaningfully.

\section{Simulation-based toy model}
\label{sec:mock}

We aim to simulate both galaxy catalogs and the Compton-$y$ map consistently.
Galaxy catalogs are often made from $N$-body simulations using a halo occupation distribution (HOD).
The process has been highly optimized in \abacushod{} \citep{AbacusHOD}.
Therefore, we choose to make a Compton parameter map from a halo catalog.

Our general strategy is to reproduce our analysis of the data with some simplification.
We make pixelated square maps from a halo catalog in a cubic periodic box using a flat-sky approximation.
We make these maps for mock Compton-$y$ measurements (true signal plus noise) and their pixel-wise standard deviation $\sigma_y$ (mimicking patterns from the overlap of ACT and DESI footprints).
We then take an HOD mock galaxy catalog produced from the same halo catalog, find the pixel corresponding to each galaxy position, and take the galaxy's tSZ SNR from that pixel.
Finally, we separate the mock galaxies into categories (bins) by this property and compute cross-correlation functions between these categories and the full set of mock galaxies.

We are aware of the advanced models for SZ-halo connection informed by hydrodynamic simulations or observations \citep[e.g.,][]{simulations-websky,baryon-pasting-algorithm,fast-baryon-painting-Liu}.
However, in this proof-of-concept work, we choose to develop a simpler model that is easier to control.

\subsection{Simple \texorpdfstring{$y$}{y} signal map from an \texorpdfstring{$N$}{N}-body simulation}
\label{sec:mock:signal}

The Compton $y$ parameter depends on the electron temperature $T_e(\bm r)$ and number density $n_e(\bm r)$ as functions of spatial position $\bm r$, whereas an $N$-body simulation provides masses and positions of discrete dark matter particles grouped in halos.
Let us recall the theoretical expression for $y$ as the integral along the line of sight \citep{Planck-SZ-map,Sunyaev-Zeldovich-1972}:
\begin{equation} \label{eq:y-parameter}
    y\qty(\bm \theta) = \int_0^{r_{\rm ls}} \frac{k_B T_e\qty(r, \bm \theta)}{m_e c^2} n_e\qty(r, \bm \theta) \sigma_T dr.
\end{equation}
The integral above is taken along the radial coordinate $r$ to the last scattering surface $r_{\rm ls}$, where CMB emerges.
$k_B$ is the Boltzmann's constant, $m_e c^2$ is the electron's rest energy and $\sigma_T$ is the Thomson cross-section.

We estimate electron temperatures $T_e$ using the motions of dark matter particles in halos.
The root-mean-square thermal velocity is analogous to the random velocity dispersion of particles.
The velocity dispersion of dark matter particles belonging to each halo is available in the \abacussummit{} \compaso{} halo catalogs \citep{CompaSO-halo-finder}.
The remaining non-trivial step is connecting dark matter to hot gas.
We assume complete ionization, primordial composition\footnote{I.e., mix of hydrogen-1 and helium-4 with the mass fraction of the latter $Y_{\rm He} = 0.2454$ \citep{Planck2018-cosmo}.}, and thermal equilibrium between ions and electrons in the gas.
We find adequate hot gas temperatures\footnote{The 3-dimensional velocity dispersion $\sigma_{3d}\sim 200~{\rm km~s^{-1}}$ yields $T_e\sim 10^6$~K.} by taking the same kinetic energy of random motions per unit mass as for dark matter particles \citep[similarly to][]{tSZ-electron-density-temperature-assumptions}:
\begin{equation} \label{eq:T_e}
    \frac12 \mu m_p \sigma_{3d}^2 \approx \frac32 k_B T_e.
\end{equation}
$\sigma_{3d}$ is the 3D velocity dispersion of dark matter particles within the halo, $k_B$ is Boltzmann's constant, and $\mu$ is the mean mass per particle in units of proton mass $m_p$, neglecting the electron mass:
\begin{equation}
    \mu \approx \qty(2 - \frac54 Y_{\rm He})^{-1},
\end{equation}
where $Y_{\rm He}$ is the mass fraction of helium.

Then, we approximately compute the number of electrons $N_e$ from the dark matter mass.
First, we obtain the baryonic matter mass using the global baryon-to-dark matter ratio given by the cosmic density parameters $\Omega_b/\Omega_{\rm cdm}$ \citep{tSZ-electron-density-temperature-assumptions}.
Second, we estimate the hot gas mass by assuming it constitutes a constant fraction of all baryons $f_{\rm hot}\approx 0.85$ \citep{cosmic-baryon-budget}.
Either of these coefficients may be imprecise, but this can be corrected for in the scaling calibration we perform in \cref{sec:mock:calibration}.
Finally, we compute the electron number $N_e$ assuming complete ionization of the hot gas and primordial composition as previously:
\begin{equation} \label{eq:N_e}
    N_e \approx \qty(1 - \frac12 Y_{\rm He}) f_{\rm hot} \frac{\Omega_b}{\Omega_{\rm cdm}} \frac{M_{\rm halo}}{m_p}.
\end{equation}

We could also have arrived at the same scaling between the number density of electrons $n_e$ and the mass density of dark matter:
\begin{equation} \label{eq:n_e}
    n_e \approx \qty(1 - \frac12 Y_{\rm He}) f_{\rm hot} \frac{\Omega_b}{\Omega_{\rm cdm}} \frac{\rho_{\rm dm}}{m_p},
\end{equation}
but scaling between densities is a stronger assumption than scaling between the total mass and number of electrons.
Moreover, we do not use the detailed density profiles for our final mock maps.

Using \cref{eq:y-parameter,eq:T_e,eq:n_e}, we can arrive at a simple expectation for (central) Compton-$y$ parameter from a halo.
First, we obtain $y \propto \sigma_{3d}^2 \int \rho_{\rm dm} dl$.
Then, we can estimate $\int \rho_{\rm dm} dl \propto M_{\rm halo}/R_{\rm halo}^2$.
A simplistic definition of a halo is a spherical region with a fixed high density\footnote{However, constant density does not hold in modern halo finders, including \compaso{} \citep{CompaSO-halo-finder}.}, so $M_{\rm halo} \propto R_{\rm halo}^3$.
Assuming virialization, $\sigma_{3d}^2 \propto M_{\rm halo}/R_{\rm halo}$.
Putting all together gives $y \propto M_{\rm halo}$, although halos can violate the underlying assumptions (and we see a steeper relation in \cref{sec:mock:Mhalo-SZ,fig:Mhalo-tSZ-relations-mocks}).

To proceed further, we use the simplifying flat-sky approximation valid for small angular scales for simplicity.
Accordingly, we switch from $\bm r = \qty(r, \bm \theta)$ decomposition to $\qty(r_\parallel, \bm r_\perp)$:
\begin{equation} \label{eq:y-parameter-flatsky}
    y\qty(\bm r_\perp) \approx \frac{k_B \sigma_T}{m_e c^2} \int_0^{r_{\max}} T_e\qty(r_\parallel, \bm r_\perp) n_e\qty(r_\parallel, \bm r_\perp) dr_\parallel.
\end{equation}

Next, we need to account for particle discreteness.
As the first rough approximation, we take infinitely narrow profiles of $T_e(\bm r) n_e(\bm r)$ for each dark matter particle with the 3-dimensional Dirac delta function $\delta^{(3)}$:
\begin{equation} 
    y\qty(\bm r_\perp) \approx \frac{k_B \sigma_T}{m_e c^2} \sum_j \int_0^{r_{\max}} T_{e, j} N_{e, j} \delta^{(3)}\qty(\bm r - \bm r_j) dr_\parallel.
\end{equation}
Here $j$ indexes dark matter particles with positions $\bm r_j$ and the corresponding electron temperature and number estimates.
We can evaluate the integral over $r_\parallel$ by decomposing the delta function in $r_\parallel$ and $r_\perp$\footnote{Namely, $\delta^{(3)}\qty(\bm r - \bm r_j) = \delta^{(1)}\qty(r_\parallel - r_{\parallel, j}) \delta^{(2)}\qty(\bm r_\perp - \bm r_{\perp, j})$. The integral of the former over $dr_\parallel$ gives 1 as long as $r_{\parallel, j}$ is not out of integration bounds.}:
\begin{equation} \label{eq:y-map-discrete-particles} 
    y\qty(\bm r_\perp) \approx \frac{k_B \sigma_T}{m_e c^2} \sum_j T_{e, j} N_{e, j} \delta^{(2)}\qty(\bm r_\perp - \bm r_{\perp, j}).
\end{equation}

In the following step, we need to discretize the Compton-$y$ map into pixels.
We can do this by averaging over pixel number $k$ with area $A_k$:
\begin{equation}
    y_k = \frac1{A_k} \int_{A_k} d\bm r_\perp y\qty(\bm r_\perp).
\end{equation}
Substituting \cref{eq:y-map-discrete-particles} into the integral above gives counts of dark matter particles\footnote{We use computational particles from a 3\% subsample in \abacus{} belonging to each halo to track the dark matter mass structure. We initially put a mass of $M_{\rm particle}/0.03$ on each of these particles. Then, if the total mass put on particles in a halo is less than the mass of the halo, we put the remaining mass at the center of this halo \citep[L2 according to][]{CompaSO-halo-finder}. If the total mass put on particles in a halo exceed the mass of the halo, we downscale the masses on particles in this halo to match the halo mass, and do not add any mass at the center of the halo.} in pixels weighted by their $T_e N_e$ (determined via \cref{eq:T_e,eq:N_e}) along with constant factors:
\begin{equation} \label{eq:y-map-pixelized}
    y_k = \frac1{A_k} \frac{k_B \sigma_T}{m_e c^2} \sum_{j:\, r_{\perp, j} \in A_k} T_{e, j} N_{e, j}.
\end{equation}

Subsequently, we need to convolve the pixelated map with the point spread function (the CMB instrument beam).
We use fine pixels at this stage.
The common beam for the ACT DR6 + {\it Planck} map has a Gaussian shape with 1.6 arcmin full width at half-maximum \citep{ACT-component-separated-maps-DR6}.
This translates to $\approx 0.9 \ihMpc$ at $z=0.8$ --- close to the typical halo size.
Thus, the exact width of the electron temperature-density profile is not so important.

Finally, we downsample the map to larger pixels, approximately matching the pixel side of 0.5 arcmin in the ACT DR6 + {\it Planck} Compton-$y$ map.
Afterwards, we also apply a 2.4 arcmin FWHM Gaussian filter analogously to our data processing.

The naive expectation $y\propto M_{\rm halo}$ for the central Compton-$y$ parameter from a halo (using \cref{eq:y-parameter,eq:T_e,eq:n_e}) can hold for halos larger than the pixel, beam and filter size.
For small, unresolved halos, there is an additional factor of halo area relative to the fixed pixel, beam or pixel area $\propto R_{\rm halo}^2 \propto M_{\rm halo}^{2/3}$.
Alternatively, \cref{eq:y-map-pixelized} can be used together with \cref{eq:T_e,eq:n_e} again.
Either way, we obtain a steeper scaling $y_{\rm map} \propto M_{\rm halo}^{5/3}$ as a simple expectation for smaller, unresolved halos.

Our simple tSZ model agrees with \cite{simple-SZ-models} in the temperature estimate from dark matter velocity dispersion (\cref{eq:T_e}) and the ratio of electron number to dark matter mass (\cref{eq:n_e,eq:N_e}, except our addition of the $f_{\rm hot}$ factor).
In the following steps, we only track cluster structure via the computational dark matter particles, justifying this by the beam smearing, pixelation and further Gaussian filtering, whereas \cite{simple-SZ-models} use more detailed density and velocity dispersion profiles.
On the other hand, their model assumes isolated halos, whereas our treatment takes into account signal accumulation from multiple systems located close on the flat sky.

\subsection{Simple noise model}
\label{sec:mock:noise}

We generate noise maps as a Gaussian random field independent of the signal for simplicity.
Signal and noise may be correlated in actual data processing (e.g., due to deprojection of different foregrounds), and this can account for some of the differences between data and mocks we notice later.
On the other hand, we incorporate realistic spatial correlations and the varying pixel-wise Compton-$y$ standard deviation from the data.
Without these, the mock clustering increases with tSZ SNR noticeably faster than in the data.

We take the power spectrum of the noise (normalized by the pixel-wise standard deviations $\sigma_y$) from a filtered noise simulation provided with the ACT DR6 products (accordingly normalized, and filtered\footnote{We applied our fiducial Gaussian filter after normalization by pixel-wise standard deviations (computed previously from 304 realizations of noise simulations). Filtering before normalization corresponds to the data processing more exactly, but it creates small-scale artifacts at the boundaries of regions observed by ACT to different depths.}).
We generate a Gaussian random field in a periodic square grid (the same as for our Compton-$y$ signal map) from this power spectrum, subtract the mean and normalize by the standard deviation.
We also create a periodic square map for the standard deviation $\sigma_y$, reproducing its distribution by area in the ACT DR6 + {\it Planck} noise maps restricted to the DESI sky footprint\footnote{In our map, $\sigma_y$ depends only on one coordinate, and increases symmetrically from the middle to the boundaries. This way, we respect the periodicity of the map inherited from the box.}.
To obtain the total signal-to-noise ratio, we add the normalized noise to the ratio of the pure signal to the standard deviation, all taken from the pixel to which a mock galaxy falls.

This treatment allows us to mimic the zero-mean Gaussian fluctuations caused by instrumental noise and some component-separation systematics.
However, it does not capture systematic biases, deviations from normal distribution and potential correlations between the signal and the noise.
As a result, negative SNR in our mocks only results from Gaussian smearing, whereas in data, systematic biases may also contribute.

\subsection{Calibration}
\label{sec:mock:calibration}

To refine our arbitrary coefficients in the $y$ signal making described above, we rescale the signal to match the 3-dimensional density of $>4\sigma$ detections (cluster candidates) in the data ($0.4<z<0.85$).
This should be good as long as there is rarely more than one significant SZ cluster per pixel.
Otherwise, we would need to match depth with the real Universe, which would be challenging with cut-sky mocks as well, because it would require bigger simulation boxes.

We match the ACT DR6 SZ cluster catalog \citep{ACT-SZ-clusters-DR6} containing 9977 confirmed objects to the 6108 peaks ($>4\sigma$) detected within the cluster search area in the ACT DR6 + {\it Planck} tSZ Compton-$y$ map after our fiducial filtering.
Because of the differences in the cluster candidate detection methods, the matching is not 1-to-1.
We restrict good matches to 3523 unique clusters within 2.4 arcmin of our peaks, among which 1709 belong to our redshift range ($0.4<z<0.85$).
Given the sky area of the ACT DR6 cluster search mask, their average density is $\approx 49 \qty(\ihGpc)^{-3}$.
From this, we extrapolate the density of our $>4\sigma$ peaks as $\approx 84 \qty(\ihGpc)^{-3}$.
We rescale the mock $y$ signal to match this density of $>4\sigma$ peaks in our first map (given that our simulation box is $\qty(2\ihGpc)^3$).
The resulting signal rescaling factor is $\approx 1.05$, which suggests that our rough assumptions in \cref{sec:mock:signal} were not far off, or their errors nearly compensated each other.

It is important to take noise into account in this calibration: without adding noise, we find the rescaling factor $\approx 1.49$.
It can be expected because even though we use a normal (symmetric) distribution for noise, higher signals are rarer\footnote{That said, true negative signals are not possible in our simplistic model, and are probably hard to achieve in reality.}.
Therefore, at a given final signal-to-noise\footnote{At least a positive value.}, more objects were shifted ``up'' by noise (have a smaller true signal) than ``down'' (from a larger true signal).

\subsection{Mock clustering}

\begin{figure}[btp]
    \centering
    \includegraphics[width=\linewidth]{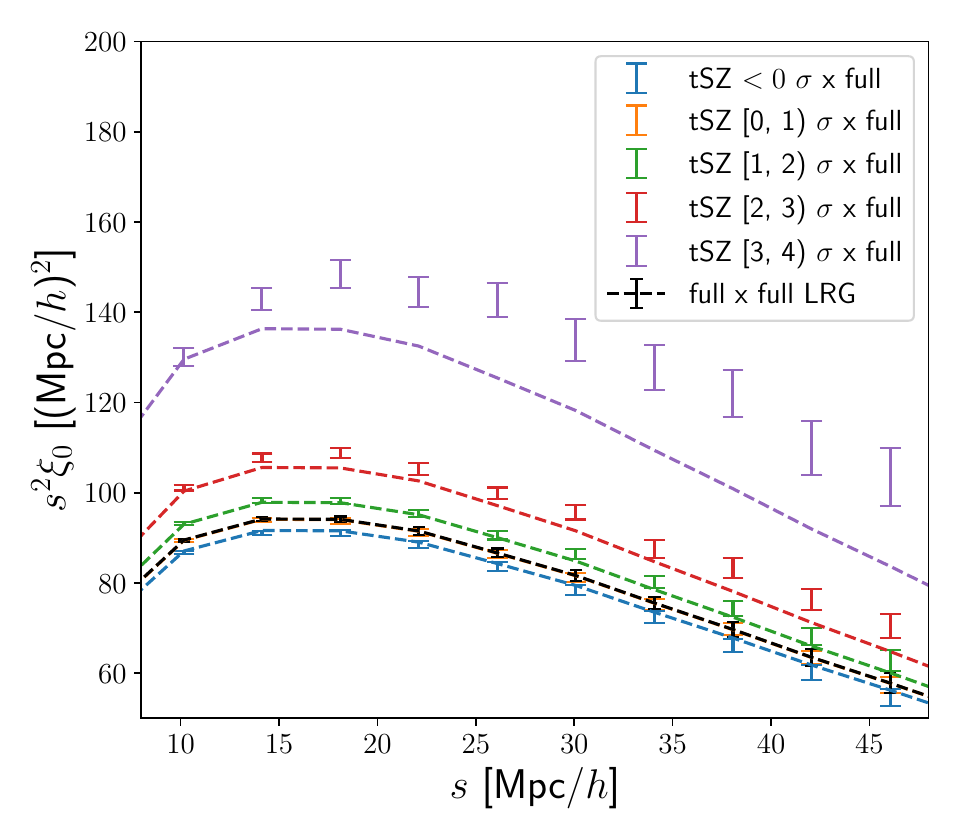}
    \caption{Mock large-scale isotropic (monopole) cross-correlation functions of different SNR bins with the full LRG sample (excluding pairs with perpendicular separation $r_p<3.4~\ihMpc$).
    Similarly to the data (\cref{fig:clustering-SNR-bins}), the clustering is enhanced at higher SNR, but less strongly.
    Also note that the $[0, 1)$ $\sigma$ bin remains very similar to the full sample.
    Colored dashed lines show the best fits obtained by scaling the full-sample autocorrelation function (black dashed line).
    The lines go below most of the points for higher-SNR bins, but keep in mind that strong correlations between the correlation function bins make fitting counterintuitive\footnote{Detailed information about all fits is included in the supplementary material: \supplementarylink{}}.
    The scaling coefficients are the relative biases shown in \cref{fig:mock-relbias-SNR-bins}.}
    \label{fig:mock-clustering-SNR-bins}
\end{figure}

\begin{figure}[htbp]
    \centering
    \includegraphics[width=\linewidth]{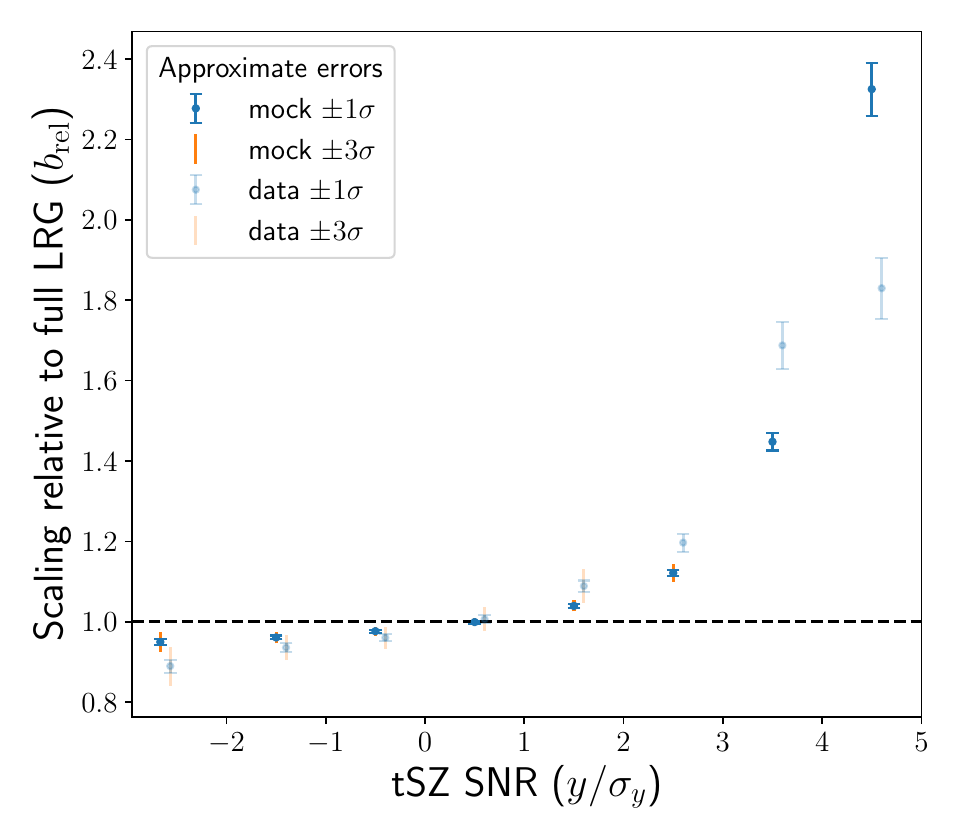}
    \caption{Mock galaxy bias of different SNR bins relative to the full LRG sample.
    Similarly to data (shaded and shifted slightly to the right for clarity), the relative bias increases with the SNR, but less rapidly near 0 and then more steeply at higher values.}
    \label{fig:mock-relbias-SNR-bins}
\end{figure}

Bringing together all the ingredients, we can finally compare the clustering of SNR bins in mocks to data.
We focus on the larger scales, showing cross-correlation function monopoles in \cref{fig:mock-clustering-SNR-bins} and relative galaxy bias in \cref{fig:mock-relbias-SNR-bins}.
We disregard the pairs with perpendicular separation $r_p<3.4~\ihMpc$, which matches the angular separation cut we used for data (0.1 degrees) at $z_{\rm mock}=0.8$.
The clustering/bias enhancement trend in mocks does not reproduce the data closely, although this should be expected from the simplicity of our modeling.
The general similarity of the trend, however, suggests that our corresponding results for the data (\cref{fig:clustering-SNR-bins,fig:relbias-SNR-bins}) are not significantly driven by systematics (which are not included in our mocks).

One of the real data systematics not modeled in our cubic mocks is fiber assignment incompleteness \citep{KP3s6-Bianchi}.
As this effect should lead to the underrepresentation of denser regions, we might expect lower relative bias for positive tSZ SNR in data compared to mocks.
However, in \cref{fig:mock-relbias-SNR-bins}, we see the opposite between 1 and $4\sigma$.
This may be partially explained by the successful mitigation of fiber assignment incompleteness effects in large-scale clustering with weights \citep{KP3s6-Bianchi,DESI2024.II.KP3} and on-sky separation ($\theta$) cut \citep{KP3s5-Pinon}, as we use both on the data side.
Correcting for fiber assignment on small scales is a much more difficult problem, and this is a significant reason why we currently do not focus on the line-of-sight clustering (analogs of \cref{fig:LoS-clustering-SNR-bins,fig:LoS-sigmas-SNR-bins}).

\subsection{Dependency on galaxy-halo connection}
\label{sec:mock:hod}

Our next step is to investigate how the different clustering in tSZ SNR bins reacts to changes in galaxy-halo connection assumptions.
Doing this with all cross-correlation functions (\cref{fig:mock-clustering-SNR-bins}) is hardly viable, because it would add even more lines to the plot, making it hard to interpret.
The relative bias plot (\cref{fig:mock-relbias-SNR-bins}) provides a cleaner way forward.

We use the 7-parameter halo occupation distribution (HOD) model (\citetalias{Zheng2007-HOD}$+f_{\rm ic}+\alpha_{\rm c}+\alpha_{\rm s}$, but with $f_{\rm ic}=1$) as in \cite{EDR_HOD_LRGQSO2023}, implemented in \abacushod{} \citep{AbacusHOD}.
5 parameters describe the expected numbers of central and satellite galaxies in each halo \citep[following][]{Zheng2007-HOD}:
\begin{align}
    \ev{N_{\rm cen}(M)} &= \frac{f_{\rm ic}}2 \qty{1+\erf\qty[\frac{\log(M/M_{\rm cut})}{\sigma}]}, \label{eq:N_cen} \\
    \ev{N_{\rm sat}(M)} &= \qty[\frac{\max(M-\kappa M_{\rm cut}, 0)}{M_1}]^\alpha \ev{N_{\rm cen}(M)}. \label{eq:N_sat}
\end{align}
$M_{\rm cut}$ sets the mass threshold for a halo to host a central galaxy.
$\sigma$ sets the rapidity of the transition from $\approx 0$ to $\approx 1$ central galaxies.
$M_1$ is a roughly typical mass for a halo hosting one satellite galaxy.
Halos below the $\kappa M_{\rm cut}$ mass threshold will not have any satellite galaxies.
$\alpha$ is the power-law index for the number of satellite galaxies.
Two remaining parameters describe the velocity bias: $\alpha_c$ for centrals by scaling the Gaussian scatter $\delta v (\sigma_{\rm LoS})$ added to the halo center of mass velocity $v_{\rm L2}$:
\begin{equation} \label{eq:alpha_c}
    v_{{\rm cen},z} = v_{{\rm L2},z} + \alpha_c \delta v (\sigma_{\rm LoS}),
\end{equation}
and $\alpha_s$ for satellites by shifting the galaxy's velocity from the velocity of a chosen dark matter computation particle $v_{{\rm p},z}$ towards the center of mass:
\begin{equation} \label{eq:alpha_s}
    v_{{\rm sat},z} = v_{{\rm L2},z} + \alpha_s (v_{{\rm p},z} - v_{{\rm L2},z}).
\end{equation}
Line-of-sight ($z$) velocities are highlighted because we compute clustering in redshift space.

\begin{figure}[tbp]
    \centering
    \includegraphics[width=\linewidth]{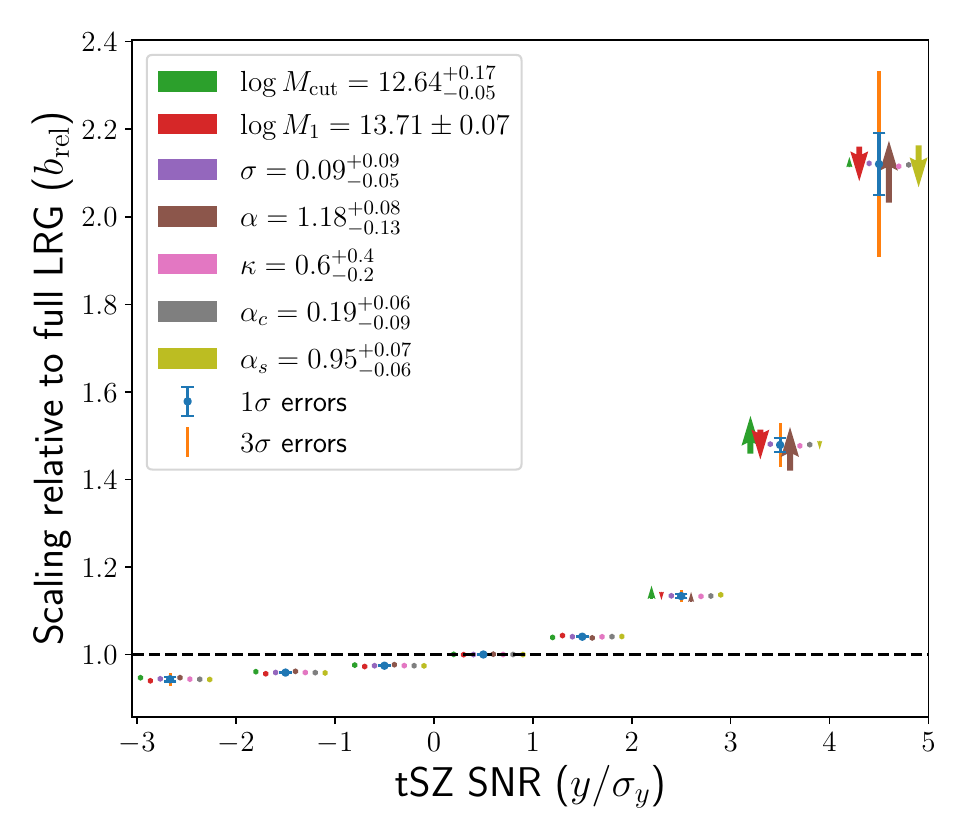}
    \includegraphics[width=\linewidth]{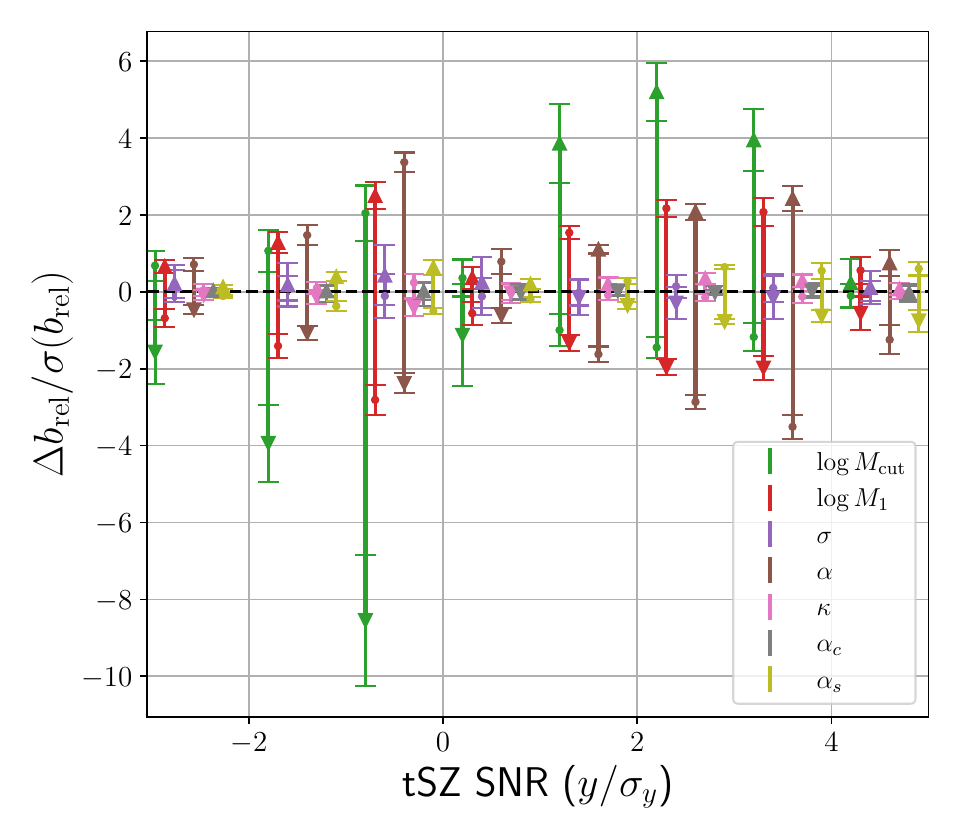}
    \caption{Effect of varying halo occupation distribution (HOD) parameters on clustering in different tSZ SNR bins.
    We shift each of the 7 parameters separately by $1\sigma$ up and down from the best fit \citep[following $0.6<z<0.8$ LRG results from][]{EDR_HOD_LRGQSO2023}, keeping the other 6 parameters fixed.
    Logarithms are decimal, and masses are in $h^{-1} M_\odot$.
    Arrows are directed from a lower value of each parameter towards the higher one.
    Top panel: absolute changes in mock galaxy bias of different SNR bins relative to the full LRG sample, averaged over 25 realizations.
    Most changes are small and hard to see.
    Bottom panel: highlighted differences in each bin's relative bias with respect to the best-fit HOD, divided by the standard deviation estimate (for the $(2 \ihGpc)^3$ box with full completeness).
    The ends of arrows represent averages over 25 realizations, the errorbars on them are for one realization (not for the mean, i.e., not divided by 5) and are estimated from the sample variance.}
    \label{fig:mock-relbias-hod-variations}
\end{figure}

We vary each of the 7 HOD parameters by $1\sigma$ up and down from the best fit, keeping the other parameters fixed.
We take the best fits and the errorbars from the \citetalias{Zheng2007-HOD}$+f_{\rm ic}+\alpha_{\rm c}+\alpha_{\rm s}$ model analysis of LRG within $0.6<z<0.8$ \citep{EDR_HOD_LRGQSO2023} in the DESI Early Data Release.
We only do not subsample for incompleteness, keeping $f_{\rm ic}=1$.
We use 25 realizations (different phases in the initial conditions) of the base \abacussummit{} $(2\ihGpc)^3$ box to obtain more robust mean shifts and estimate the sample covariance.
For each realization, we make tSZ signal and noise maps\footnote{As described in \cref{sec:mock:signal,sec:mock:noise}. We continue using the SNR rescaling calibrated on the first ({\tt ph000}) realization in \cref{sec:mock:calibration}.}, save galaxy catalogs for the additional 14 HOD variants, split them into tSZ SNR bins according to the Compton-$y$ SNR map (the same across HOD variants for one realization), compute the clustering estimators and fit the relative scaling (bias).

We keep the same tSZ map with varying HOD parameters, so the same halos remain in each SNR bin.
$M_{\rm cut}$ and $\sigma$ change the number of LRG centrals and satellites in these halos.
$M_1$, $\alpha$ and $\kappa$ only affect the number of satellites.
$\alpha_c$ and $\alpha_s$ only alter redshift-space positions of centrals and satellites respectively.
All these affect the redshift-space clustering of LRGs.

We show the changes in relative biases in each bin due to these HOD variations in the top panel of \cref{fig:mock-relbias-hod-variations}.
Most of the differences are hard to discern (except some in the $3-4$ and $4-5$ SNR bins), but the errorbars are also hard to see in many of these cases.
To address this issue, we take the differences between HOD variants and the best-fit HOD and divide by the errorbar estimate (for the best-fit HOD).
The results are in the bottom panel of \cref{fig:mock-relbias-hod-variations}.
$[0, 1)~\sigma_y$ bin shows some of the least significant shifts, meaning that its bias relative to the full sample remains close to 1.
Among the small differences for other bins, we can see larger effects of varying $M_{\rm cut}$, $M_1$ and $\alpha$.

\begin{table}[btp]
    \centering
    \begin{tabular}{|c|c|c|}
    \hline
    $(\Delta n_{\rm gal})/n_{\rm gal}$ & $-1\sigma$ & $+1\sigma$ \\
    \hline
    $\log M_{\rm cut} = 12.64^{+0.17}_{-0.05}$ & $13.2\%$ & $-34.8\%$ \\
	\hline
	$\log M_1 = 13.71\pm 0.07$ & $2.9\%$ & $-2.4\%$ \\
	\hline
	$\sigma = 0.09^{+0.09}_{-0.05}$ & $-1.8\%$ & $7.1\%$ \\
	\hline
	$\alpha = 1.18^{+0.08}_{-0.13}$ & $1.6\%$ & $-0.7\%$ \\
	\hline
	$\kappa = 0.6^{+0.4}_{-0.2}$ & $1.2\%$ & $-2.2\%$ \\
    \hline
    $\alpha_c = 0.19^{+0.06}_{-0.09}$ & \multicolumn{2}{c|}{0} \\
    \hline
    $\alpha_s = 0.95^{+0.07}_{-0.06}$ & \multicolumn{2}{c|}{0} \\
    \hline
    \end{tabular}
    \caption{Relative changes in mock number density with varying halo occupation distribution (HOD) parameters.
    We shift each of the 7 parameters separately by $1\sigma$ up and down from the best fit \citep[following $0.6<z<0.8$ LRG results from][]{EDR_HOD_LRGQSO2023}, keeping the other 6 parameters fixed.
    Logarithms are decimal, and masses are in $h^{-1} M_\odot$.
    These values are averaged over 25 realizations.
    Note that we are keeping full completeness ($f_{\rm ic}=1$ in \cref{eq:N_cen,eq:N_sat}).
    The velocity bias parameters $\alpha_c$ and $\alpha_s$ (\cref{eq:alpha_c,eq:alpha_s}) do not change the number density by construction.}
    \label{tab:mock-density-hod-variations}
\end{table}

The relative bias shifts change sign between the $[0,1)~\sigma$ bin and the $[1,2)~\sigma$ bin for all parameters except $\alpha_c$\footnote{For $\alpha_c$, the sign flips between $[-1,0)~\sigma$ and $[0,1)~\sigma$, but all the differences have very low significance.} in the bottom panel of \cref{fig:mock-relbias-hod-variations}.
The direction flip in itself is only natural, because the average relative bias (weighted by the number\footnote{For data, it should be weighted by the sum of clustering weights. But in our cubic mocks, all the galaxies have weights set to 1.} of galaxies in each corresponding SNR bin) should be 1\footnote{As it is defined as the scaling of the full-sample auto-correlation function to best match the SNR bin-full sample cross-correlation functions}.
However, it is not clear to us why the change of sign is around $1\sigma$ tSZ detection significance specifically for most of the HOD parameters.

Despite several significant differences in the bottom panel of \cref{fig:mock-relbias-hod-variations}, the relative bias pattern seems relatively insensitive to HOD.
First, the variations due to $M_{\rm cut}$, $M_1$ and $\alpha$ are directed similarly and may be hard to disentangle.
Second, the scale of HOD parameter variations is set by the precision of small-scale clustering in DESI Early Data Release (having a small volume but high completeness), whereas the errorbars of our relative biases are computed for a $(2 \ihGpc)^3$ box (with full completeness).
Third, we varied each parameter individually by its 1-dimensional standard deviation up and down instead of following the degeneracy directions.
This should produce higher-significance differences in the full-sample clustering and number density.
We show the relative changes in number density in \cref{tab:mock-density-hod-variations}, and $M_{\rm cut}$ variations have the largest effect (however, note that we are keeping full completeness, $f_{\rm ic}=1$ in \cref{eq:N_cen,eq:N_sat}).
The stability of the relative bias pattern with HOD would be beneficial for the precision and robustness of cosmological inference from tSZ-split clustering, and we will validate this preliminary finding further in future work.

\subsection{Relating halo mass to thermal Sunyaev-Zeldovich observables}
\label{sec:mock:Mhalo-SZ}

Simulations also allow us to access the true halo properties, unlike the data.
With this, we can explore connections with SZ properties (bearing in mind they are produced approximately).

\begin{figure}[tbp]
    \centering
    \includegraphics[width=\linewidth]{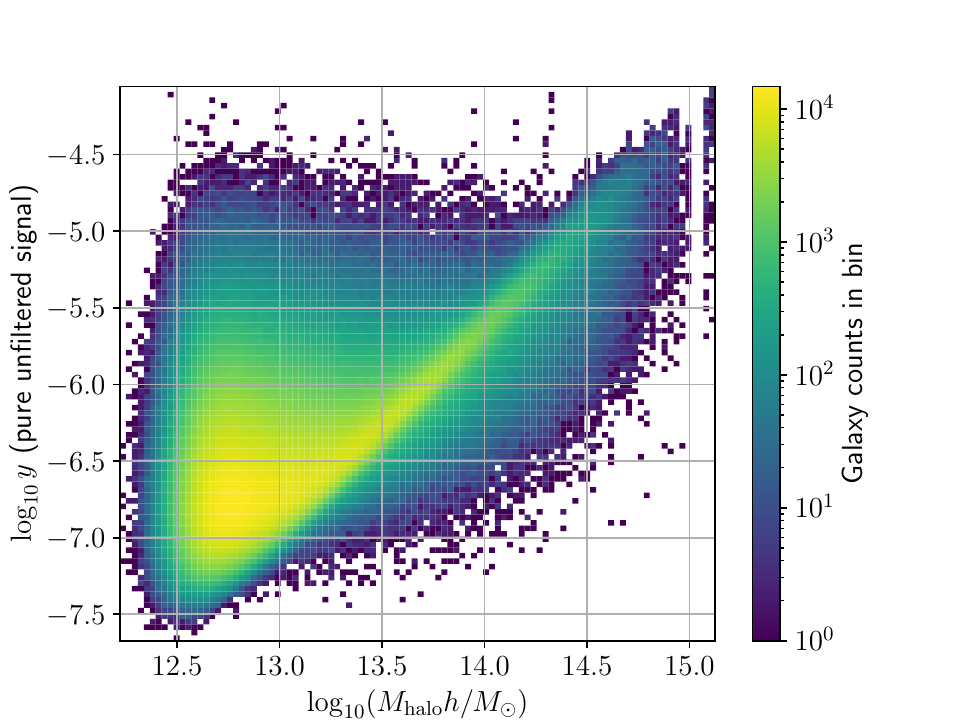}
    \includegraphics[width=\linewidth]{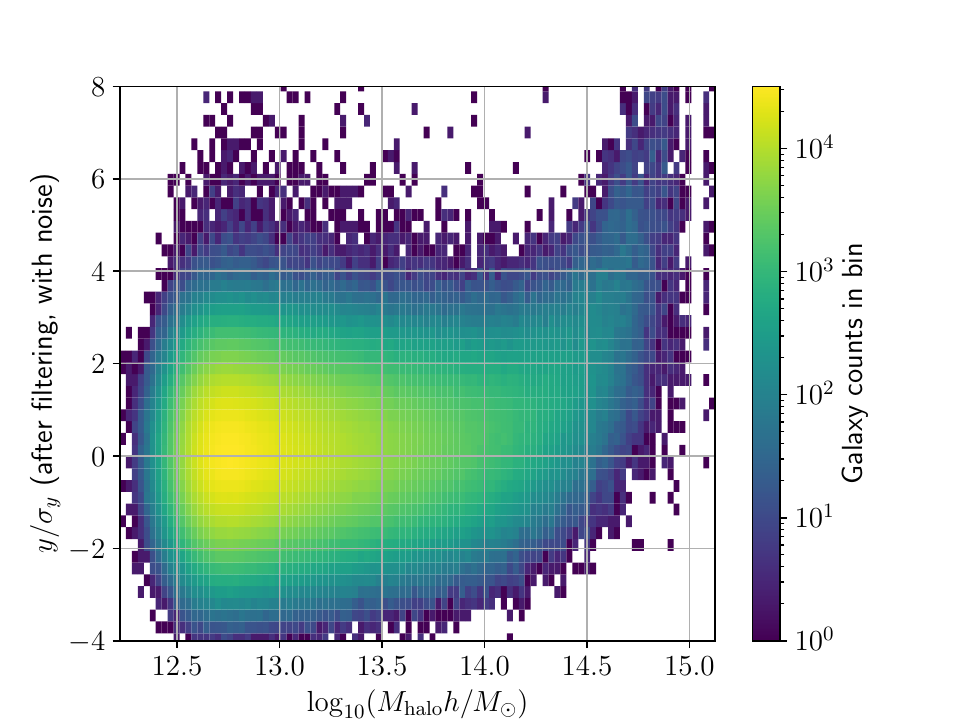}
    \includegraphics[width=\linewidth]{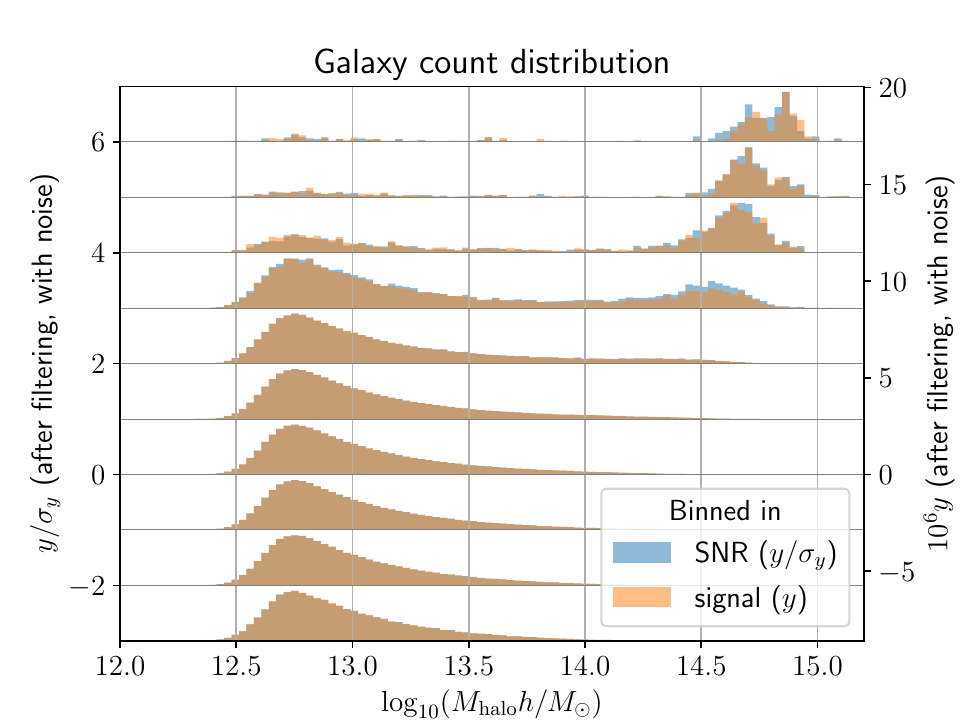}
    \caption{Histograms displaying correlations between LRG host halo masses and tSZ quantities in the corresponding pixels in our mock map
    \citep[assuming ACT DR6 + {\it Planck} characteristics following][]{ACT-component-separated-maps-DR6,ACT-noise-simulations-DR6}.
    Top: 2D histogram using the noiseless $y$ parameter without filtering.
    Middle: 2D histogram using the tSZ SNR with our fiducial filter and simplistic noise.
    Bottom: 1D histograms of halo mass in SNR (blue) or signal (orange) bins (the two distributions are very similar; both observables include our fiducial filter and simplistic noise).
    All histograms are weighted by galaxy counts from one mock catalog with the fiducial HOD model (see \cref{sec:mock:hod}).}
    \label{fig:Mhalo-tSZ-relations-mocks}
\end{figure}

We present correlations between host halo mass and several tSZ quantities for mock galaxies\footnote{We include these properties for all galaxies in the supplementary material, \supplementarylink{}, and encourage searching for other interesting relations.} in \cref{fig:Mhalo-tSZ-relations-mocks}.
We count LRGs from a single mock galaxy catalog created with the fiducial HOD model, having their host halo mass listed in the mock catalog\footnote{Therefore, the absence of $M_{\rm halo} \lesssim 10^{12.3}~M_\odot/h$ is not a limitation of halos in the \abacussummit{} simulation but a reflection of the best-fit HOD model from \cite{EDR_HOD_LRGQSO2023}. We tried plotting halo mass distributions, but found it much less informative due to a very strong contribution of smaller halos down to $\sim 10^{11}~M_\odot/h$ that would almost never host LRGs. The absence of $M_{\rm halo} \lesssim 10^{11}~M_\odot/h$ halos in \abacussummit{} "base" boxes we used might lead to misestimation of the mock tSZ signal, but their contribution is likely not very significant (as we naively expect $y_{\rm map} \propto M_{\rm halo}^{5/3}$ for each very low-mass halo from \cref{sec:mock:signal}, and the more realistic power index may be even steeper).}, and obtain tSZ quantities by matching them to our map pixels.
In the upper panel of \cref{fig:Mhalo-tSZ-relations-mocks} we see a power-law relation between the halo mass and the unfiltered noiseless Compton-$y$.
The power-law looks slightly steeper than $y \propto M_{\rm halo}$ (estimated with big simplifications in \cref{sec:mock:signal}; the index may be similar to $\approx 1.08$ from Eq.~(4) in \citealt{ACT-SZ-clusters-DR6}).
The trend smears up for smaller halo masses; this may be caused by such halos appearing close (within the point spread function) to slightly larger halos.
The middle panel of \cref{fig:Mhalo-tSZ-relations-mocks} shows a range of SNR values centered on zero for lower halo masses (as expected from Gaussian noise and negligible signal) and a noticeable increase for larger halo masses.

In the bottom panel of \cref{fig:Mhalo-tSZ-relations-mocks}, we explore the realistic observables in more detail by first splitting the data into SNR bins and then producing 1D histograms of halo mass in each.
The lower-SNR bins ($y/\sigma_y<4$) are dominated by galaxies from $M_{\rm halo} \approx 10^{12.7} M_\odot$ halos, but the distributions have a long higher-mass tail, and it extends further and further with increasing SNR.
In the $[3, 4)~\sigma_y$ bin, a secondary bump at the high mass ends becomes visible, and it dominates in the higher ($y/\sigma_y\ge 4$) bins.
Thus, a qualitative transition occurs between 3 and 4 $\sigma_y$.

We also repeat the binning in the signal (Compton-$y$ parameter not divided by $\sigma_y$).
We have chosen the bin width as the mean $\sigma_y$ across the map for easier comparison with the SNR binning.
The two resulting distributions in the bottom panel of \cref{fig:Mhalo-tSZ-relations-mocks} are very similar (differences are visible only in the higher-SNR bins), reinforcing our arguments from \cref{sec:measurements:methodology}.

A well-known or calibrated relation between the galaxy host halo masses and tSZ observables can be used to make theoretical predictions about tSZ-split galaxy clustering, especially on larger scales.
Furthermore, if this relation is less dependent on cosmology than halo mass-density relations, it would enable more robust cosmological inference beyond the 2-point function.
A theoretical or semi-analytical model would be a very attractive alternative to a simulation-based model \citep[like for density-split clustering in][]{density-split-clustering-sim-based-model} requiring a large number of high-quality mocks.

\section{Conclusions}
\label{sec:conclusions}

We propose to split the spectroscopic galaxy samples into distinct subpopulations by the signal-to-noise ratio for the thermal Sunyaev-Zeldovich effect.
This effect has long been associated with galaxy clusters \citep{Sunyaev-Zeldovich-1970} and is an established tool for their detection and property determination \citep{Planck-SZ-clusters,ACT-SZ-clusters-DR5,ACT-SZ-clusters-mass-calibration-DR5,SPT-clusters-w-DES+HST-WL,SPT-SZ-clusters-2025,ACT-SZ-clusters-DR6}.
However, the rigorously detected clusters \citep[$>4\sigma$ in][]{ACT-SZ-clusters-DR5,ACT-SZ-clusters-DR6} are relatively uncommon.
We focus on the larger areas of the map with lower signal-to-noise, and demonstrate that they can enable us to extract more valuable cosmological information from galaxy clustering.

From large-scale clustering (\cref{fig:clustering-SNR-bins}), we find that a higher tSZ signal-to-noise ratio corresponds to galaxies with a higher bias (\cref{fig:relbias-SNR-bins}).
We see a similar increase in clustering amplitude in the projected correlation function (\cref{fig:projected-clustering-SNR-bins}), which can be used with larger photometric galaxy samples \citep[e.g.,][]{LRG-samples-for-cross-correlations}.
The clustering enhancement with tSZ can allow us to select luminous red galaxy sub-samples and test the consistency of large-scale patterns in them \citep[e.g., following][]{LSS-color-dependent-stochasticity},
or better understand their biases to obtain tighter constraints on primordial non-Gaussianity \citep[in the spirit of][]{multi-tracer-PNG-forecasts}.

From small-scale line-of-sight clustering, we find that the velocity dispersion increases considerably with tSZ SNR (\cref{fig:LoS-sigmas-SNR-bins}).
This is likely a clean indication of strong non-perturbative non-linearities (Fingers of God) which can be removed to improve the theoretical modeling \citep{removing-FoG}.
We further show how the velocity dispersion is reduced by putting an upper threshold on the tSZ SNR (\cref{fig:LoS-sigmas-cumulative-SNR-bins}).

We also see that galaxies with higher tSZ SNR have higher numbers of close neighbors (\cref{fig:avg-no-close-neighbors-cylinders-SNR-bins}).
We remind the reader that the number of close on-sky neighbors in DESI data is suppressed by fiber collisions \citep{DESI2024.II.KP3,KP3s6-Bianchi}.
As an external indicator, tSZ could help inform the galaxy multiplet studies using spectroscopic \citep[e.g.,][]{DESI-galaxy-multiplets-tidal-field} or photometric data.

We have built a simple mock tSZ map that allowed us to reproduce general trends from the data (\cref{fig:mock-clustering-SNR-bins,fig:mock-relbias-SNR-bins}).
However, we note differences, and investigating their causes will be our next priority.
We also tentatively investigate how the galaxy-halo connection (halo occupation distribution) affects the relative bias of different tSZ bins (\cref{fig:mock-relbias-hod-variations}) and find relatively small changes.

In further work, we aim to develop cosmological inference from tSZ-split clustering (e.g., constraining the cosmic growth of structure).
One avenue is to build a simulation-based model or inference pipeline, like for density-split \citep{density-split-clustering-sim-based-model} or density-marked clustering \citep{density-marked-PS-SBI}.
However, this requires investing in a large number of accurate simulations with different cosmological and galaxy-halo connection parameters.
We have presented preliminary simulation-based results on the halo mass-tSZ observables relation in \cref{fig:Mhalo-tSZ-relations-mocks}, which look sensible and consistent with previous works.
Together with hints of weak sensitivity to the galaxy-halo connection, this gives us hope of building a semi-analytical model by calibrating the relation between the halo mass and the thermal Sunyaev-Zeldovich observables using relatively few simulations.

tSZ splitting could be useful beyond luminous red galaxies, although it might be less efficient.
At higher redshifts, there are fewer SZ clusters \citep{ACT-SZ-clusters-DR6}, and the density of galaxies is lower, meaning that splitting would be more harmful for the signal-to-noise ratio of the measurements.
At lower redshifts \citep[like DESI Bright Galaxy Sample/Survey,][]{BGS.TS.Hahn.2023}, X-ray observations may be a cleaner indicator of clusters, and the higher density allows for a better direct detection of close multiplets in spectroscopic data.

Combining 3D galaxy surveys and Sunyaev-Zeldovich data is very promising.
As well as the number of spectra with DESI, the SZ measurements are improving rapidly with the current and next CMB experiments like Simons Observatory \citep[SO,][]{SO} and CMB-S4 \citep{CMBS4,CMBS4white}.
The overlap of the SO large-aperture telescope (LAT) survey with DESI is similar to ACT DR6, but the combined depth is expected to improve from 10~$\mu$K-arcmin in ACT DR6 \citep{ACT-component-separated-maps-DR6} to 6~$\mu$K-arcmin for the baseline SO LAT \citep{SO} and 2.4~$\mu$K-arcmin for the enhanced SO LAT expected to be complete by 2028 \citep{SO-LAT-enhanced}.
There are also possibilities to use larger photometric samples further in celestial South from DESI Legacy Imaging surveys \citep[e.g., extended LRG sample][]{LRG-samples-for-cross-correlations} and LSST \citep{LSST} in the future with projected statistics or using photometric redshifts.
Joint analysis of different measurements has high potential to improve our measurements and uncover new tensions.

\section*{Acknowledgments}

We thank Antón Baleato Lizancos and Ravi K. Sheth for their very fruitful feedback and suggestions.
MR and DJE have been supported by U.S. Department of Energy grant DE-SC0013718 and by the Simons Foundation Investigator program.

This material is based upon work supported by the U.S. Department of Energy (DOE), Office of Science, Office of High-Energy Physics, under Contract No. DE–AC02–05CH11231, and by the National Energy Research Scientific Computing Center, a DOE Office of Science User Facility under the same contract. Additional support for DESI was provided by the U.S. National Science Foundation (NSF), Division of Astronomical Sciences under Contract No. AST-0950945 to the NSF’s National Optical-Infrared Astronomy Research Laboratory; the Science and Technology Facilities Council of the United Kingdom; the Gordon and Betty Moore Foundation; the Heising-Simons Foundation; the French Alternative Energies and Atomic Energy Commission (CEA); the National Council of Humanities, Science and Technology of Mexico (CONAHCYT); the Ministry of Science, Innovation and Universities of Spain (MICIU/AEI/10.13039/501100011033), and by the DESI Member Institutions: \url{https://www.desi.lbl.gov/collaborating-institutions}. Any opinions, findings, and conclusions or recommendations expressed in this material are those of the author(s) and do not necessarily reflect the views of the U. S. National Science Foundation, the U. S. Department of Energy, or any of the listed funding agencies.

The authors are honored to be permitted to conduct scientific research on I'oligam Du'ag (Kitt Peak), a mountain with particular significance to the Tohono O’odham Nation.

This work has used the following software packages: \textsc{astropy} \citep{astropy:2013, astropy:2018, astropy:2022}, \textsc{healpy} and \textsc{HEALPix}\footnote{\url{https://healpix.sourceforge.net}} \citep{Zonca2019, 2005ApJ...622..759G,healpy_8404216}, \textsc{Jupyter} \citep{2007CSE.....9c..21P, kluyver2016jupyter}, \textsc{matplotlib} \citep{Hunter:2007}, \textsc{numpy} \citep{numpy}, \textsc{pixell}\footnote{\url{https://github.com/simonsobs/pixell}}, \pycorr{}\footnote{\url{https://github.com/cosmodesi/pycorr}} \citep{pycorr,corrfunc-1,corrfunc-2}, \textsc{python} \citep{python}, \textsc{scipy} \citep{2020SciPy-NMeth, scipy_10155614}, and \textsc{scikit-learn} \citep{scikit-learn, sklearn_api, scikit-learn_10034229}.

This research has used NASA's Astrophysics Data System.
Software citation information has been aggregated using \texttt{\href{https://www.tomwagg.com/software-citation-station/}{The Software Citation Station}} \citep{software-citation-station-paper, software-citation-station-zenodo}.

\section*{Data Availability}

DESI Data Release 1 is available at \url{https://data.desi.lbl.gov/doc/releases/dr1/}; the clustering catalogs used in this work can be found at \url{https://data.desi.lbl.gov/public/dr1/survey/catalogs/dr1/LSS/iron/LSScats/v1.5/}.
ACT Data Release 6 maps and noise simulations used in this work are available at \url{https://portal.nersc.gov/project/act/dr6_nilc/}.
All the points from the figures, together with some additional data and the source code, are available at \supplementarylink{}.


\bibliographystyle{mnras}
\bibliography{DESI_supporting_papers,references,software}




\end{document}